\documentclass[runningheads]{llncs}
\usepackage[T1]{fontenc}
\usepackage{graphicx}
\RequirePackage[ruled,linesnumbered,noend]{algorithm2e}

\SetCommentSty{mycommfont}
\DontPrintSemicolon
\RequirePackage[dvipsnames]{xcolor}

\usepackage{adjustbox} % allows adjusting of boxes
\usepackage{amssymb}
\usepackage{amsmath}
\usepackage[toc,page]{appendix} % support appendices
\usepackage{array}
\usepackage{booktabs}
\usepackage{caption}
\usepackage{csquotes} % handles quotes automatically
\usepackage{graphicx}
\usepackage{hyperref} % allows autoref/hyperlinks/etc.
\usepackage{cleveref}
\crefname{subsection}{Sec.}{Sec.}
\crefname{section}{Sec.}{Sec.}
\crefname{subsubsection}{Sec.}{Sec.}
\crefname{chapter}{Ch.}{Ch.}
\crefname{part}{Part}{Parts}
\crefname{figure}{Fig.}{Fig.}
\crefname{table}{Tab.}{Tab.}
\crefname{definition}{Def.}{Def.}
\crefname{enumi}{}{}
\crefname{algocf}{Alg.}{Alg.}
\Crefname{algocf}{Alg.}{Alg.}
\crefname{algocfline}{line}{lines}
\crefname{equation}{Equation}{Equations}
\crefname{algorithm}{Alg.}{Alg.}

\usepackage{blkarray}
\usepackage{mathtools}
\usepackage{setspace}
\usepackage{subcaption}
\usepackage{tabularx}
\usepackage{upgreek}
\usepackage{wrapfig}

\usepackage{lineno}
\usepackage{lipsum}
\begin{document}
\title{Computing Alignments for Partially-ordered Traces Through Petri Net Unfoldings}
\titlerunning{Partial Order-Based Alignments via Petri net Unfolding}
\author{Ariba Siddiqui\inst{1} \and
Wil M. P. van der Aalst\inst{1}\orcidID{0000-0002-0955-6940} \and
Daniel Schuster\inst{1}\orcidID{0000-0002-6512-9580}}
\authorrunning{A.\ Siddiqui et al.}
\institute{
RWTH Aachen University, Aachen, Germany\\
\email{ariba.siddiqui@rwth-aachen.de, wvdaalst@pads.rwth-aachen.de, schuster@pads.rwth-aachen.de}}
\maketitle              % typeset the header of the contribution
\begin{abstract}
%Business processes are essential for maintaining efficiency and compliance across various industries. These processes are typically represented using models like BPMN and Petri nets, but in reality, deviations due to compliance issues, technical errors, or business changes often occur.
Conformance checking techniques aim to provide diagnostics on the conformity between process models and event data. Conventional methods, such as trace
alignments, assume strict total ordering of events, leading to inaccuracies when timestamps
are overlapping, coarse, or missing. 
In contrast, existing methods that support partially ordered events rely upon the interleaving semantics of Petri
nets, the reachability graphs, which suffer from the state space explosion problem.
%Besides, this view also forces the solution to create a partially ordered alignment almost artificially. 
This paper proposes an improved approach to conformance checking based
upon partially ordered event data by utilizing Petri net unfolding, which leverages partial-order semantics of Petri nets to represent concurrency and uncertainty in
event logs more effectively. Unlike existing methods, our approach offers a streamlined one-step solution, improving efficiency in the computation of alignments. Additionally, we introduce a
novel visualization technique for partially ordered unfolding-based alignments. We implement unfolding-based alignments with its user-friendly insights in a conformance analysis tool. Our experimental evaluation, conducted on synthetic and real-world event logs, demonstrates that the unfolding-based approach is particularly robust in handling high degrees of parallelism and complexity in process models.

\end{abstract}

\section{Introduction}

\textit{Conformance checking}, a major task within process mining, deals with comparing event data with process models, typically represented using notations like BPMN \cite{bpmn} or Petri Nets \cite{petri-nets}. \textit{Trace alignments}~\cite{van2012replaying} are considered to be state-of-the-art in conformance checking. These techniques aim to search for alignments with minimum deviations, with Adriansyah et al.\ \cite{van2012replaying} introducing a method using a shortest-path search using $A^*$ \cite{astar} in a Synchronous Product Net (SPN) of the trace net and the model net. Another method \cite{complete_interpretable} relies on a unified representation of process models and event logs based on \textit{event structures}, a well-known model of concurrency. The method returns a set of statements in natural language describing behavior allowed by the model but not observed in the log and vice versa.

One problem with trace alignment algorithms is that they assume totally-ordered event data and produce totally-ordered alignments. In a more recent work by Lu et al.\ \cite{lu2015conformance}, this limitation is overcome. A partially-ordered trace is converted to an occurrence net (i.e., a simple Petri net without choices), and an SPN is computed between the occurrence net and the normative model. From hereon, the shortest path computation is done on the state space of the product net (just like \cite{van2012replaying}) using the $A^*$ algorithm and a predefined cost function. As a firing sequence with minimum cost is found, it is replayed on the net while unfolding it into a new net. That is, while replaying, whenever a place is seen for the second time (owing to a loop), it is cloned into a new place. The result is an occurrence net, which can be converted into a partially-ordered alignment (\textit{p-alignment}). This results in a two-step computation solely because of the reliance on interleaving semantics of Petri nets, the \textit{reachability graphs}, which additionally suffer from the well-known state space explosion \cite{StateSpaceProbValmari}, limiting its scalability to larger instances of problems with high concurrency.

To address the issue of state space explosion in the interleaving semantics of Petri nets, researchers in model checking \cite{esparza1994model}, diagnosis \cite{benveniste2003diagnosis}, and planning \cite{bonet2008directed,hickmott2007planning} have extensively explored the Petri net unfolding process. This partial-order semantics, introduced in \cite{nielsen1981petri} and further detailed as \textit{branching processes} in \cite{engelfriet1991branching}, offers an alternative to interleaving semantics. A branching process unfolds all partially-ordered runs of a Petri net into a tree-like structure, representing concurrency, causality, and choice points. Although comprehensive, these processes are often infinite. McMillan's seminal work \cite{mcmillan1993using} introduced a method to construct a finite prefix of the branching process that retains state reachability information. This prefix, which models concurrency more compactly than the reachability graphs \cite{reisig2016understanding}, is typically much smaller. However, McMillan's algorithm can produce unnecessarily large prefixes. To address this, Esparza et al.\ proposed the \textit{ERV unfolding algorithm} \cite{esparza1996improvement}, which generalizes McMillan's algorithm using \textit{adequate orders}. Since these prefixes encode all information on reachable states of a Petri net, they can effectively be used for shortest-path computation problems. Esparza et al. showed that the reachability problem is NP-complete relative to prefix size \cite{esparza2001unfolding}, with their \textit{OnTheFly} $ERV$ variant proving most efficient amongst all unfolding-based reachability techniques. 

Despite advances in alignment computation, partial order-based conformance checking, and Petri net unfolding, these domains have yet to converge. 
This paper addresses this gap by applying Petri net unfolding to optimal alignment computation based on partially ordered event data. 
We apply Petri net unfolding to replace reachability graphs to compute \textit{ unfolding-based alignments}, using an algorithm we term $ERV[\triangleleft_c]$. 
Our method offers a straightforward, one-step solution, avoiding the need to create a partially ordered alignment from execution sequences artificially.
A key advantage is faster computation times, particularly when there is high concurrency between events. As a second contribution, we propose a method for visualizing the results of unfolding-based alignments using a technique from \cite{schuster2021visualizing}. This method visualizes trace variants from partially ordered event data in chevron-based views and is well-suited for graph-based alignments. Unlike existing methods, our approach makes the relationship between log and model more explicit. We evaluate our algorithm using synthetic event data with varying levels of parallelism, showing that unfolding-based alignments perform better under increasing parallelism. Regarding runtime, the (directed) unfolding-based approach outperforms the classic method by Lu et al.\ \cite{lu2015conformance}. However, in lower parallelism settings, our method faces higher computational costs. Further experiments with real-world data confirm these findings, demonstrating that unfolding-based algorithms are more robust as model complexity grows.

\section{Preliminaries}

In process mining, an \textit{event log} captures the occurrence of activities within a business process execution. Logs are analyzed to discover process models, check conformance, or optimize performance. Each event includes \textit{start} and \textit{end} timestamps, a \textit{case identifier}, and a label for the event. Case identifiers group events into a single \textit{case} representing a process execution. Conformance checking compares each case with the normative process model, and deviations are analyzed across the log. A key challenge is representing temporal relationships, especially for concurrently occurring events. To address this, we define \textit{partially-ordered traces (p-traces)} to represent concurrent activities the need for total ordering.

% In process mining, an \textit{event log} consists of \textit{events} that capture the occurrence of activities within a business process execution. These logs are typically analyzed to discover process models, check conformance, or optimize performance. Each event in the event log includes two timestamps indicating its \textit{start} and \textit{end}, along with a \textit{case identifier} and a \textit{label} representing the specific event. The case identifiers are used to group multiple events into a single \textit{case}, representing an individual process execution.  Conformance checking is typically performed by comparing each case execution with the discovered or normative process model. Deviations found can be analyzed and further aggregated across the entire event log. A crucial challenge in analyzing event logs is representing the temporal relationships between events, especially when events occur concurrently. Therefore, we define \textit{partially-ordered traces (p-traces)}, which allow us to represent concurrent activities within a single case without the need for total ordering.

\begin{definition}[Partially-ordered Trace (P-Trace)]
\label{def:p-trace}
For an event log $E$, in which all events share a common case identifier $c$, a partially-ordered trace is a labeled directed acyclic graph $\rho_c=((E, \prec_{\rho_c}), l_{\rho_c})$ where $E$ is a finite set of graph nodes and $\prec_{\rho_c}$ is a partial order over the nodes of the graph such that for any two events $a$ and $b$, $a \prec_{\rho_c} b$ iff the end timestamp of $a$ is strictly less than the start timestamp of $b$. $l_{\rho_c}: E \rightarrow \mathcal{L}$ is a labeling function for the nodes, where $\mathcal{L}$ is the universe of events labels of E. Each edge in a p-trace is referred to as a dependency between two events $e_i$ and $e_j$, indicating that $e_i$ has led to the execution of $e_j$.
\end{definition}

\begin{figure}[h!]
    \centering
    \includegraphics[width=\linewidth]{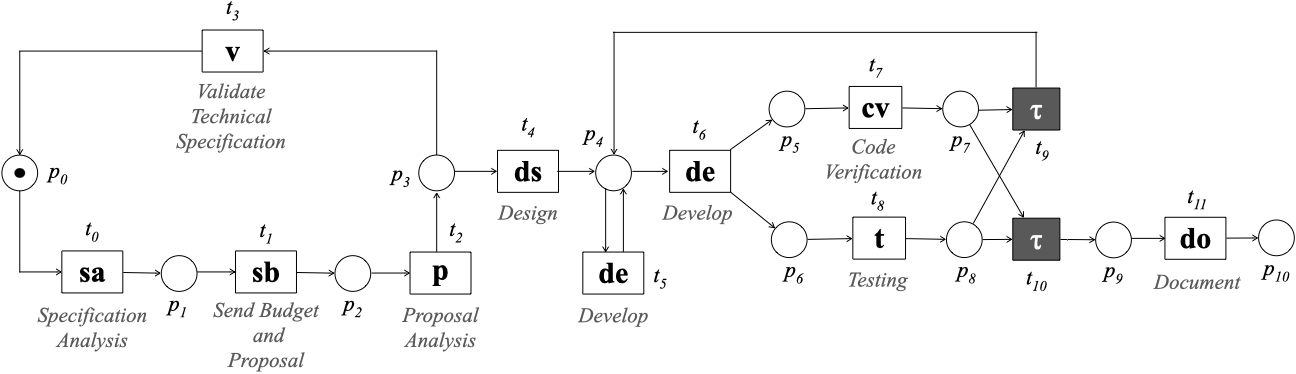}
    \caption{An example Petri net $\mathcal{N}_{mgmt}$ of a software development management process with the initial marking $M_{init} = [p_0]$ and final marking $M_{final}=[p_{10}]$}
    \label{fig:pn-mgmt}
\end{figure}

In this paper, we focus on Petri nets as the process modelling formalism. The notations of Petri nets used are largely based upon \cite{carmona2018conformance}. In its general, simplest form, Petri nets can be seen as graphs that consist of
\textit{places}, \textit{transitions}, and directed edges between them. Places (depicted as circles) and transitions (depicted as squares) represent the two types of nodes of the graph. An example Petri net $\mathcal{N}_{mgmt}$ is shown in \Cref{fig:pn-mgmt}, modeling a software project management process with 11 places and 12 transitions. 
% The process starts with `Specification Analysis`, followed by `Send Budget and Proposal`, where the client reviews the proposal. If necessary, technical specifications are validated before the process either restarts or moves to `Design`. The development phase (`Develop`) occurs iteratively, and if bugs are found, development is revisited. Once completed, the process moves to `Documentation` and terminates. Each transition represents a task, with labels like `sa` for `Specification Analysis` and `sb` for `Send Budget and Proposal`. The event `Develop` is represented by two transitions labeled `de`. A silent transition, labeled `$\uptau$`, is used for routing or missing tasks and does not correspond to any event. We define the event labels as $L^{\uptau} = L \cup \{\uptau\}$, where $\uptau$ is not in the original set $\mathcal{L}$ of event labels.

\begin{definition}[Labeled Petri Net, System Net]
    A \textit{labeled Petri net} (in short, \textit{net}) is a bipartite graph $\mathcal{N}=(P,T,F,\lambda)$ with a disjoint finite set of nodes consisting of \textit{places} $P$ and \textit{transitions} $T$, a set of (directed) arcs or \textit{flow relations} $F \subseteq ((P \times T) \cup (T \times P))$, and a labeling function $\lambda: T \rightarrow L^{\uptau}$ assigning an event label or $\uptau$ (label of a silent transition) to each transition. A \textit{system net} $SN = (\mathcal{N}, M_{init}, M_{final})$ is a labeled Petri net with a well-defined initial marking $M_{init}$ and final marking $M_{final}$.
\end{definition}

\begin{definition}[Pre- and Postset]
    Let $\mathcal{N}=(P,T,F,\lambda)$ be a labeled Petri net and $x \in P \cup T$ be a node of $\mathcal{N}$.
    The set \( \pre x = \{ y \in P \cup T \mid (y,x) \in F \} \) is called the \textit{preset}, and \( \post x = \{ y \in P \cup T \mid (x,y) \in F \} \) is called the \textit{postset} of $x$.
\end{definition}

We make the following assumptions: the sets of places are always non-empty (i.e., $P \neq \emptyset$), transitions always have incoming and outgoing edges, and the pre- and postsets of transitions are always finite. 

To model the dynamic properties of a net, the notion of \textit{marking} exists. It is a multiset over places $P$ of a net, in which the multiplicity of a place denotes the number of \textit{tokens} assigned to it. %Graphically, tokens are drawn as black dots in the place nodes. 
The net in \Cref{fig:pn-mgmt} has an initial marking $m_0=[p_0]$. A marking is said to be \textit{reachable} if there exists a sequence of transition \textit{firings} from the initial state of a net, where a transition firing follows the well-known firing semantics of Petri nets \cite{reisig2016understanding}.
In this paper, we assume all nets to be 1-safe, meaning no place contains more than one token at a time.

The dynamic behavior of Petri nets can be viewed in two ways: the first, called \textit{interleaving semantics}, considers firing sequences as possible execution paths. The second, \textit{true concurrency semantics}, views execution as partially-ordered sequences, representing runs (hereby referred to as \textit{distributed runs}) as partial orders. These partial orders can be modeled with simpler structure nets called \textit{occurrence nets}, obtained by `unfolding' system nets into a tree-like structure. Occurrence nets represent partially-ordered event sequences from the initial marking, together forming \textit{branching processes}. Stopping the unfolding at different instances yields different processes, but \textit{the} unique \textit{unfolding} is the complete branching process obtained by unfolding as much as possible. The subsequent definitions in this section are largely based upon \cite{esparza1996improvement}. Prior to introducing occurrence nets and causal nets, we define the notions of \textit{causality} and \textit{conflict} as follows---1. two nodes $x$ and $y$ are in a \textit{causal relation} with each other if there exists a path between $x$ to $y$, 2. two nodes are in \textit{conflict} with each other if there exist two different paths to these nodes which start at the same point and diverge immediately afterwards (although later on they can converge again).

\begin{definition}[Occurrence Net, Causal Net]
Let $\mathcal{N} = (B, E, F, \lambda)$ be a labeled Petri net. We have an unlabeled net $\mathcal{O}=(B, E, F)$ consisting of the transitions, places and flow relation of $\mathcal{N}$:
\begin{enumerate}
    \item $\mathcal{O}$ is an \textit{occurrence net}, if the following conditions hold:
    \begin{enumerate}
        \item $\forall b \in B (|\pre b| \leq 1)$, no place in $\mathcal{O}$ contains two or more incoming edges;
        \item $\nexists e \in E$ st. it is in \textit{conflict} with itself;
        \item $F^{+} \cap (F^{-1})^+ = \emptyset$, $\mathcal{O}$ is acyclic;
        \item $\mathcal{O}$ is finitely preceded, i.e., for every $x \in B \cup E$, the set of elements $y \in B \cup E$ such that there exists a path from $y$ to $x$ is finite.
    \end{enumerate}
    Places and transitions in an occurrence net are called as \textit{conditions} $B$ and \textit{events} $E$. 
    \item $\mathcal{O}$ is a \textit{causal net}, if (a)-(d) hold, in addition to:
    \begin{enumerate}
        \item[(e)] $\forall b \in B(|\post b | \leq 1)$, no place in $\mathcal{O}$ contains two or more outgoing edges
    \end{enumerate}
\end{enumerate}
\end{definition}

Those labeled occurrence nets obtained by unfolding system nets are referred to as \textit{branching processes}, formalized as below:\\

\begin{definition}[Branching Process]
    A \textit{branching process} of a system net $SN = ((P,T,F,\allowbreak \lambda), M_{init}, M_{final})$ is a labeled occurrence net $\upbeta=(\mathcal{O},h)=(B, E, F, h)$ where the labeling function $h: B \cup E \rightarrow P \cup T$ is a \textit{net homomorphism} from $\upbeta$ to $SN$ such that:
     $\forall v,v' \in B \cup E. (\pre v = \pre v' \wedge h(v)=h(v') \implies v = v').$ It is simply a mapping between nets that preserves the nature of nodes and the environment of nodes. It associates the conditions/events of a branching process to the places/transitions of its system net. For a formal definition of a net homomorphism, we refer to \cite{romer2000theorie}.
\end{definition}

\begin{figure}[tb]
    \centering
    \includegraphics[width=0.75\linewidth]{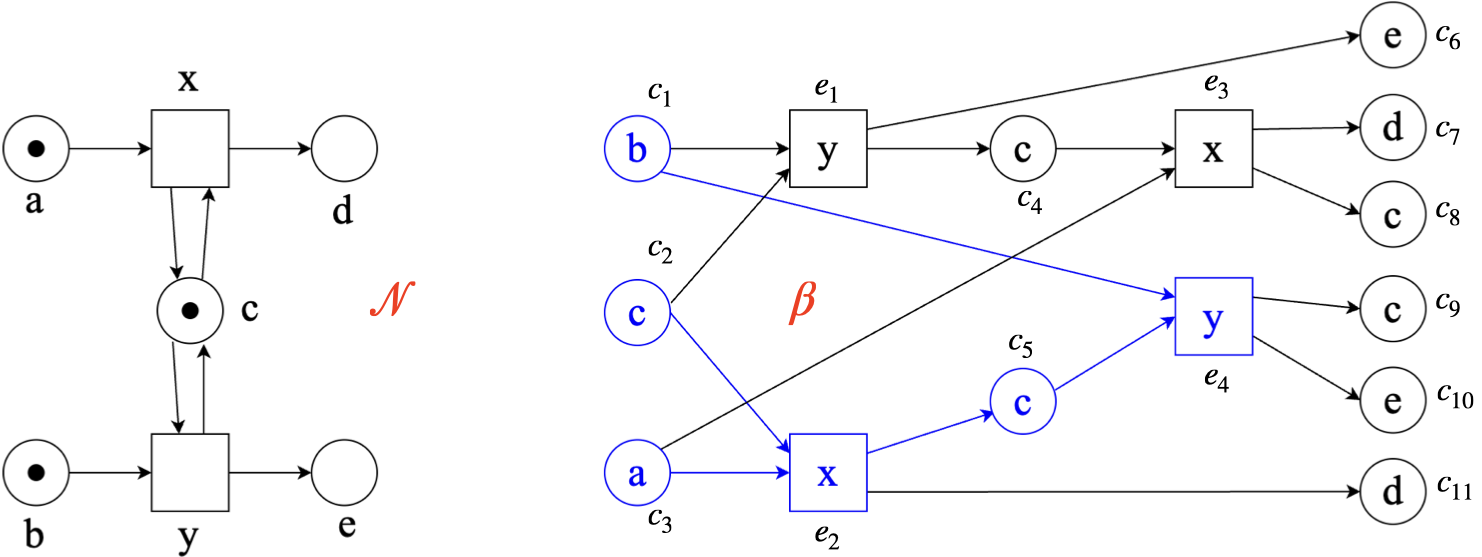}
    \caption{A system net $\mathcal{N}$ (left) with $M_{init}=[a,b,c]$, $M_{final}=[d,c,e]$, and its branching processes $\beta$ (right), which is also a complete finite prefix of the unfolding of $\mathcal{N}$.}
    \label{fig:n1-branching-processes}
\end{figure}

\Cref{fig:n1-branching-processes} shows the branching process $\beta$ of a Petri net $\mathcal{N}$. It consists of two different distributed runs corresponding to firing sequences of $x$ followed by $y$ (highlighted in blue) and $y$ followed by $x$, respectively. Labels inside each node of $\beta$ represent the places/transitions it maps to. Notice how no condition (place) in the highlighted run has more than one incoming/outgoing arc, and all events are related in a causal or concurrency relationship. Therefore, it is a causal net. Another thing to notice in the highlighted distributed run of $\beta$ is that when the run includes conditions $c_9$ and $c_{10}$, it has no outgoing arcs that enable any more transitions in the Petri net. Such a distributed run is called \textit{complete}.

The unfolding process of a system net starts with the initial conditions and extends it iteratively with the new possible events and a condition for every output place of the newly added event. This process can typically be infinite. McMillan's \cite{mcmillan1993using} (and ERV \cite{esparza1996improvement}) algorithm builds a \textit{complete finite prefix} of the unfolding, which terminates when the unfolding represents all of the reachable markings of the Petri net. The foundations of the algorithm are laid by the key concepts of \textit{configurations} and \textit{cuts}. A finite configuration $C$ represents the set of events that have occurred so far in a distributed run. For the highlighted run in $\beta$ (cf. \Cref{fig:n1-branching-processes}), $\{e_2,e_4\}$ form a finite configuration while $\{e_1,e_2\}$ does not, since $e_1$ and $e_2$ are in conflict with each other. A configuration is related to a marking $Mark(C)$ by identifying which conditions will contain a token after firing all events in configuration $C$. Thus, $Mark(C)$ corresponds to a final marking in the original system net, reached by firing all transitions labeled by the events in a given finite configuration. When a finite configuration is associated with an event, it is termed as a \textit{local configuration}.

\begin{definition}[Local Configuration]
\label{def:local-configuration}
    For an occurrence net $\mathcal{O}=(B, E, F)$, a \textit{local configuration} $[e]$ of an event $e \in E$ is defined by $\{e' \in E \ \mid \ (e',e) \in F^*\}$
\end{definition}

 It represents a minimal set of all the events denoted by $[e]$ that must have been fired for the event $e$ to be enabled. For example, in $\upbeta$, $[e_4]=\{e_2,e_4\}$. Each local configuration is also associated with the distributed run it represents. It consists of precisely the events in $[e]$, in addition to the conditions and arcs connecting them. We term it as a \textit{local distributed run}.

\begin{definition}[Local Distributed Run]
\label{def:local-dist-run}
    Let $\mathcal{O}=(B, E, F)$ be an occurrence net. A \textit{local distributed run} of an event $e \in E$ is a causal net $\mathcal{O}_e=(B_e, E_e, F_e)$ where:
    \begin{itemize}
        \item $B_e = \{c \in B \ \mid \ (c,e) \in F^* \}$ is the set of conditions,
        \item $E_e = [e]$ is the set of events, and
        \item $F_e$ is the flow relation such that $F_e(x,y) = F(x,y), \forall x,y \in B_e \cup C_e$.
    \end{itemize}
\end{definition}

The key to building a complete finite prefix (due to McMillan) is to identify those events where the unfolding process can be stopped without any loss of information. Those events are called \textit{cut-off events} and can be defined in terms of \textit{adequate order} on configurations \cite{esparza1996improvement,mcmillan1993using,chatain2007well}. Adequate orders are simply strict partial orders to choose between the extensions of two events when the $Mark$ of their respective local configurations is the same, and one of them precedes the other one with respect to the partial order. The one not chosen to be extended will thus be the \textit{cut-off event}. For a strict partial order to qualify as an adequate order, it must satisfy certain properties (cf. \cite{esparza1996improvement} for details). Thus, the unfolding can be ceased as soon as an event $e$ takes a prefix to a marking which has already been induced by a previously added event $e'$ such that $[e'] \triangleleft [e]$. The event $e$ is marked as a cutoff event. Assume an adequate order $\triangleleft$ where $e_1 \triangleleft e_2$ iff $|[e_1]|<|[e_2]|$ and a current state of marking as $\{d,c,e\}$ in $\mathcal{N}$ of \Cref{fig:n1-branching-processes}. If there were a transition $z$ from place $d$ to $a$ in $\mathcal{N}$, the event mapped to $z$ in $\beta$ would have been marked as a cut-off event. This is because it would bring the marking of $\mathcal{N}$ to $\{a,c,e\}$, already represented in $\beta$ by the local distributed run of $e_1$, for which $|[e_1]|$ is lesser already.

\section{Computation of Unfolding-Based Alignments}

\begin{figure}[tb]
    \centering
    \includegraphics[width=.8\linewidth]{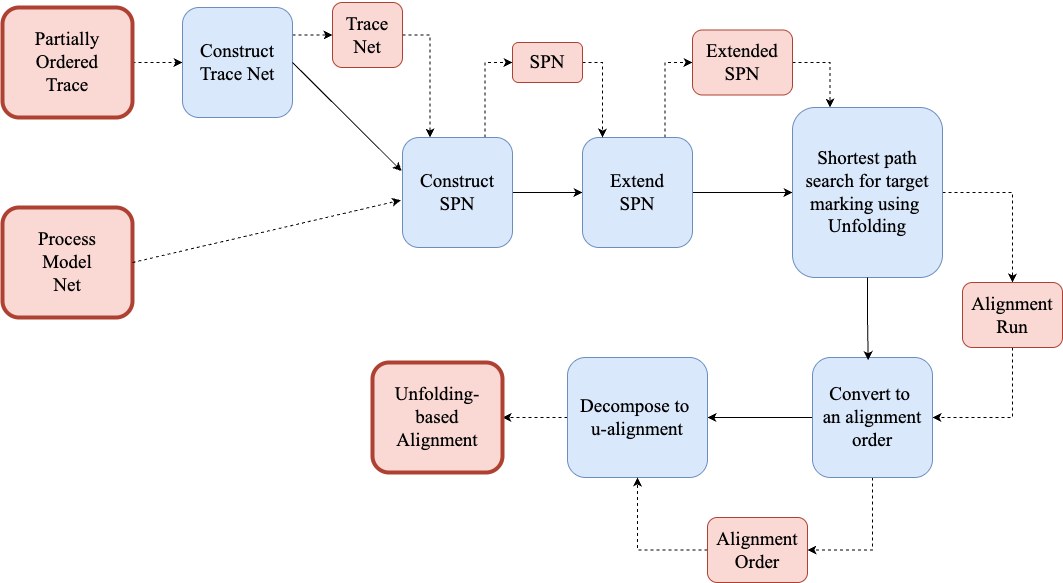}
    \caption{Overview of the proposed method to compute unfolding-based alignments
given a partially-ordered trace and a process model. Input and output of each step are
shown through dashed arrows. Various steps are highlighted in blue}
    \label{fig:pipeline-method}
\end{figure}

We describe our proposed approach as illustrated in \Cref{fig:pipeline-method}. Given a partially ordered trace (p-trace) and a model net, we first model the p-trace as a \textit{trace net} \cite{lu2015conformance}, a Petri net capturing the p-trace's behavior. Each event becomes a transition, and places represent dependencies between events.
%Input places are added for events without predecessors, and output places for events without successors. 
Next, we construct the SPN (cf. \Cref{def:sync-prod-net} in Appendix) with the model net. The product is constructed by pairing transitions in one net with transitions in the other net that have the same label [51]. This helps to explicitly model all possible alignment moves. Then we extend it by adding a target transition and place, and revise the final marking to the target place. Using unfolding algorithms $ERV[\triangleleft_c]$ or $ERV[\triangleleft_h]$ (cf. \Cref{sec:directed-unf,sec:erv-unfolding-algo}), we unfold the extended SPN to find a minimum-cost distributed run leading to the target transition, yielding an optimal alignment run. Finally, we use this run to build an alignment order and decompose it into \textit{unfolding-based alignments (u-alignments)}. The goal of these alignments is to relate p-trace with the closest run through the model net (one with minimum deviations) in a meaningful and interpretable way.

We begin by defining u-alignments, in the context of an alignment function that relates synchronous moves when the alignment is decomposed into its log and model parts. This refinement allows for a clearer relationship between log and model visualizations.

\begin{definition}[Unfolding-based Alignment (u-alignment)]
\label{def:main-definition}
Let $\rho_c=((E, \allowbreak \prec_{\rho_c}),l_{\rho_c})$ denote a p-trace and $\mathcal{N} = ((P,T,F,\lambda), \allowbreak M_{init}, M_{final})$ be a system net with induced partial order $G_{\mathcal{N}}=((T',\prec_{\mathcal{N}}),l_{\mathcal{N}})$ over a run of $\mathcal{N}$ (we refer to it as a \textit{model run}), where $T'$ is an arbitrary set such that $T' \cap E = \emptyset$ and $l_{\mathcal{N}}:T' \rightarrow rng(\lambda)$ is a labeling function. An \textit{unfolding-based alignment (u-alignment)} is a tuple $(\rho_c, G_{\mathcal{N}}, \upvarphi)$ where $\upvarphi : E' \rightarrow T'$ is an injective function, we refer to as the \textit{alignment function} defined over a set $E' \subseteq E$ and $T' \subseteq T$ such that:

\begin{enumerate}
    \item $\forall e \in E', t \in T'$ st. $( l_{\rho_c}(e) = l_\mathcal{N}(t)) \Leftrightarrow \upvarphi(e) = t$ (nodes with same activity labels are mapped together; they are synchronous), and
    \item A directed graph consisting of nodes and edges from the p-trace $\rho_c$, model run $G_{\mathcal{N}}$ such that there is a single node for each pair of synchronous nodes is a strict partial order. $\rho_c \biguplus G_\mathcal{N}$ denotes a partial order that combines $\rho_c$ and $G_\mathcal{N}$ in such a way. We refer to it as an \textit{alignment order}.
\end{enumerate}

\end{definition}

\begin{figure}[ht]
    \centering
    \begin{subfigure}{\textwidth}
         \centering
         \includegraphics[width=\textwidth]{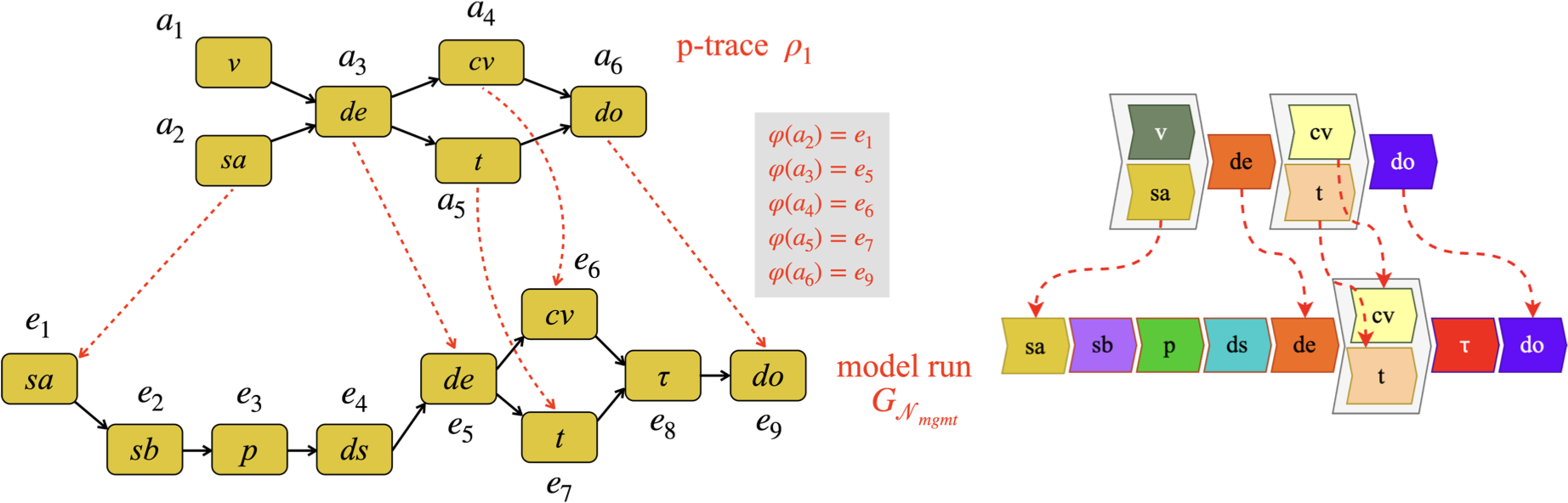}
         \caption{A u-alignment comprising of an arbitrary p-trace $\rho_1$ and a partial order $G_{\mathcal{N}_{mgmt}}$ induced by an example model run through $\mathcal{N}_{mgmt}$ \Cref{fig:pn-mgmt}. Mapping by the alignment function $\upvarphi$ is shown in dashed red arrows. A chevron-based view of the alignment is visualized on the right}
         \label{fig:alignment-function}
     \end{subfigure}
     \begin{subfigure}{\textwidth}
         \centering
         \includegraphics[width=0.55\textwidth]{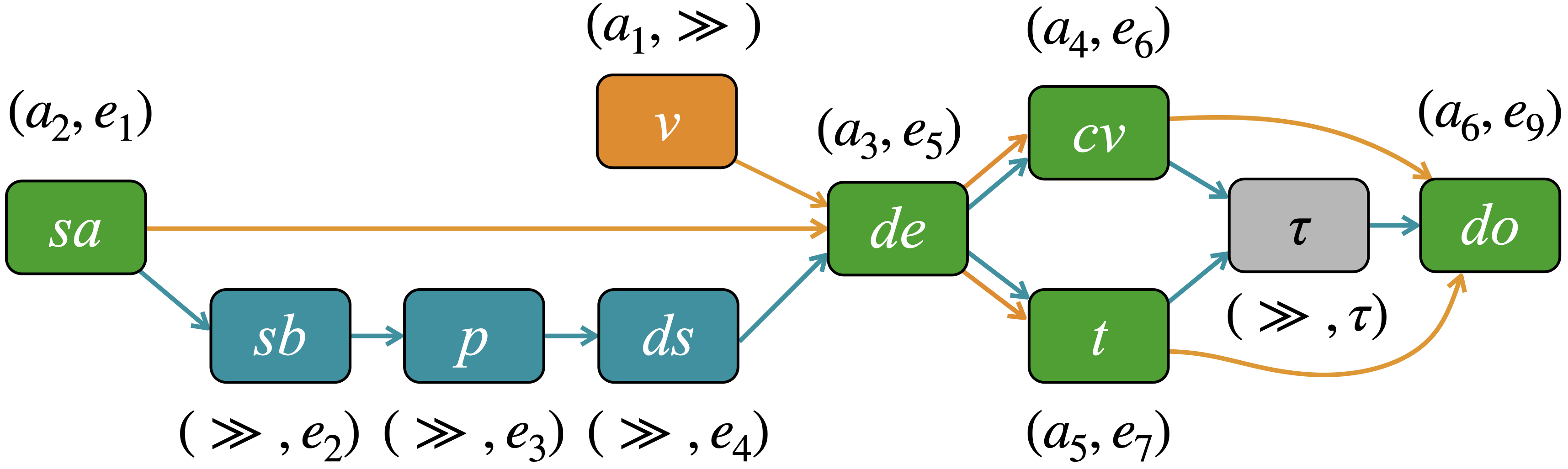}
         \caption{An alignment order $\rho_1 \biguplus G_{\mathcal{N}_{mgmt}}$. The partial order contains a single node $(a_i,e_j)$ for each pair of synchronous nodes, colored in green. All other nodes and dependencies are colored according to their origin — orange if it originates from $\rho_1$ (a node $(a_i,\gg)$), blue if from $G_{\mathcal{N}_{mgmt}}$ (a node $(\gg, e_j)$). A grey node $(\gg, \uptau)$ represents an invisible move originating from model} 
         \label{fig:alignment-order}
     \end{subfigure}
    \captionsetup[subfigure]{justification=justified, width=\textwidth}
    \caption{An example of a u-alignment along with its corresponding alignment order}
    \label{fig:main-definition}
\end{figure}

\Cref{fig:alignment-function} shows an illustration of how in a u-alignment, an alignment function maps a p-trace $\rho_1$ to a partial order induced by an example model run $G_{\mathcal{N}_{mgmt}}$ through $\mathcal{N}_{mgmt}$. The alignment function $\upvarphi$ is defined over nodes of p-trace $\{a_2,a_3,a_4,a_5,a_6\}$ such that $\upvarphi(a_2)=e_1$, $\upvarphi(a_3)=e_5$, $\upvarphi(a_4)=e_6$, $\upvarphi(a_5)=e_7$ and $\upvarphi(a_6)=e_9$. Note that $\upvarphi$ maps only nodes with the same labels; the label for $a_2$ and $e_1$ is `sa', and so on. \Cref{fig:alignment-order} shows an alignment order for $\rho_1$ and $G_{\mathcal{N}_{mgmt}}$, a partially ordered structure of \textit{alignment moves}. Nodes in this graph are colored according to where they originate from. An orange node represents a \textit{log move} from the p-trace, while a blue node represents a \textit{model move} from the model run. A green node indicates a \textit{synchronous move}, originating from both the p-trace and the model run. A grey node denotes an invisible move from the model run. Edges represent dependencies, categorized as \textit{model dependencies} (blue) or \textit{log dependencies} (orange), based on their origin. Given such an alignment, the partial orders being directed acyclic graphs, can be visualized in a chevron-based view based upon recursive sequential and parallel partitioning of the graphs, a visualization technique from \cite{schuster2021visualizing}. The result of the same is illustrated on the right of \Cref{fig:alignment-function}. Notice how this view retains the relationship between the log and model parts through the alignment function.

By understanding how to combine a trace with a model run to create an alignment order, we can similarly decompose an alignment order into two partial orders while defining the alignment function simultaneously. This yields a u-alignment.
% These partial orders are isomorphic to $\rho_c$ and an arbitrary run over the model, respectively
Given an alignment order $G_\upvarphi=((M, \prec_{\upvarphi}), l_{\upvarphi})$, we denote the two decomposed partial orders as $G_{\upvarphi \downarrow 1}$ and $G_{\upvarphi \downarrow 2}$, respectively, i.e., $G_\upvarphi \rightarrow G_{\upvarphi \downarrow 1} \biguplus G_{\upvarphi \downarrow 2}$. An alignment function $\upvarphi$ is defined for every synchronous move. For our example in \Cref{fig:main-definition}, decomposing the alignment order from \Cref{fig:alignment-order} results in two partial orders isomorphic to $\rho_1$ and $G_{\mathcal{N}_{mgmt}}$ in \Cref{fig:alignment-function}. 

In order to compute u-alignments from an SPN, it is important to realize that each transition of an SPN of a trace net and a model net represents an alignment move \cite{adriansyah2013alignment}. Thus, distributed runs of the product net define a partially-ordered structure of alignment moves, an alignment order. Consequently, the problem of computing alignment(s) for a given p-trace and a model net can be reformulated as the problem of finding complete distributed run(s) in its SPN. Since we seek an \textit{optimal} run corresponding to the fewest log and model moves, we use a \textit{default cost function} \cite{adriansyah2013memory}, which assigns higher costs to log and model moves than to synchronous and invisible moves, reducing the problem to finding minimum-cost complete distributed runs in the product net.

\subsection{Finding Optimal Distributed Runs}

We already know that a complete finite prefix of a system net encodes all its reachable markings. We exploit this to search for a target marking corresponding to the final marking of an SPN in order to find complete distributed run(s). To enable search for a target marking in an SPN, we apply the standard encoding trick of adding a \textit{target transition} $t^*$ to the product net, which consumes the final marking. To preserve 1-safeness, we also add a target place $p^*$ as a successor to $t^*$. \\

% \begin{figure}[!ht]
% \centering
% \begin{subfigure}[t]{0.19\textwidth}
%      \centering
%      \includegraphics[width=\textwidth]{figures/p-trace-running-eg.png}
%      \caption{A p-trace}
%      \label{fig:p-trace-running-eg}
%  \end{subfigure}
%  \hspace*{1cm}
%  \begin{subfigure}[t]{0.33\textwidth}
%      \centering
%      \includegraphics[width=\textwidth]{figures/model-net-running-eg.png}
%      \caption{An easy sound model net with $M_{init}=[p_0]$ and $M_{final}=[p_2]$}
%      \label{fig:model-net-running-eg}
%  \end{subfigure}
%  \vfill
%  \begin{subfigure}[b]{\textwidth}
%      \centering
%      \includegraphics[width=.63\textwidth]{figures/encoding-trick.png}
%      \caption{An extended synchronous product net $SN_{\otimes}^{ext}$ of an SPN of a trace net (induced by the p-trace in \Cref{fig:p-trace-running-eg}) and the model net in \Cref{fig:model-net-running-eg}. The target transition is $t^*$, and a target place $p^*$ is a successor to $t^*$. For this system net, $M_{init,ext}=[p_{(\blacktriangleright,b)},p_0]$ and $M_{final,ext}=[p^*]$}
%      \label{fig:extended-spn}
%  \end{subfigure}
% \caption{An extended synchronous product net} %of a trace net (induced by a p-trace) with a model net}
% \label{fig:encoding-trick}
% \end{figure}

\begin{figure}[tb]
    \centering
    \includegraphics[width=0.85\linewidth]{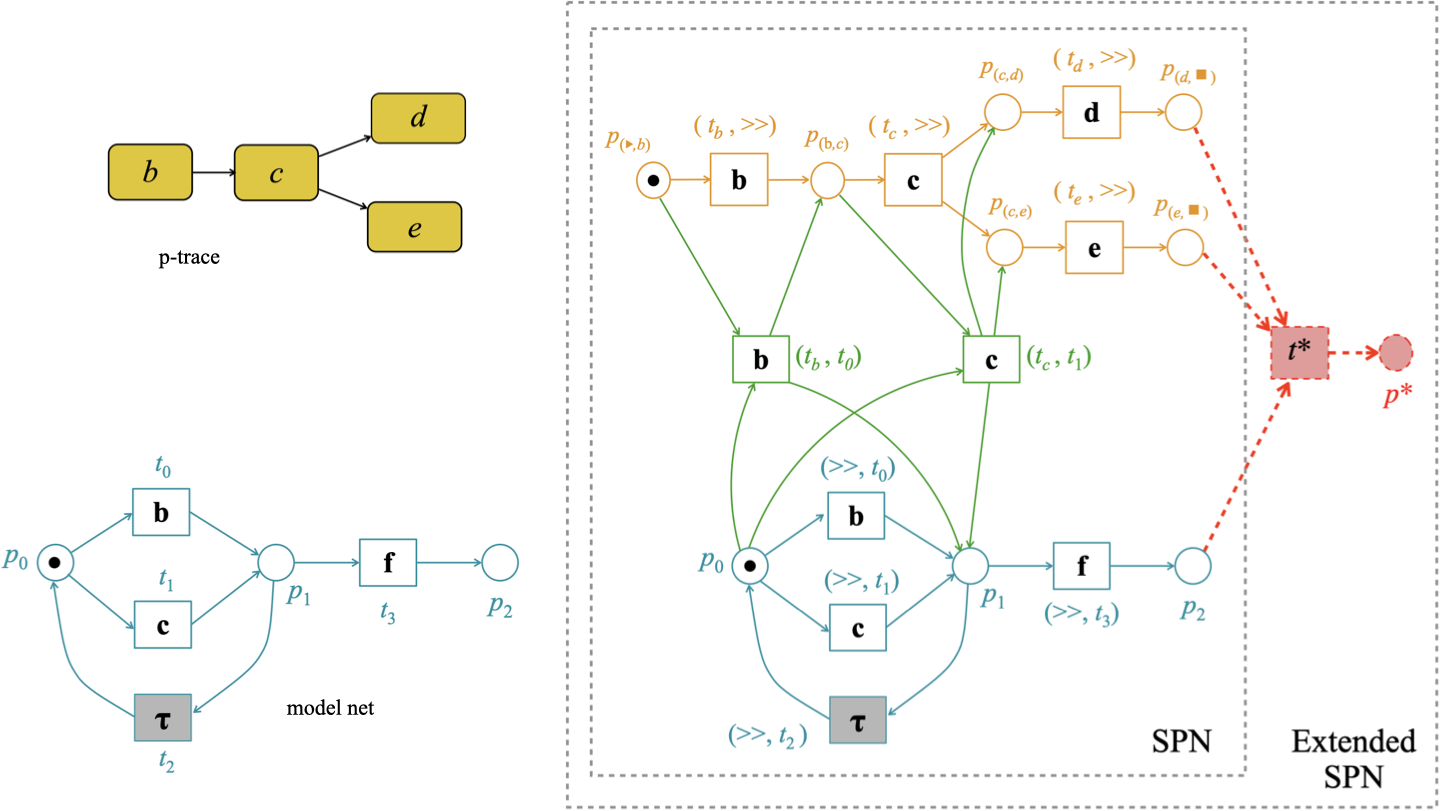}
    \caption{An extended synchronous product net $SN_{\otimes}^{ext}$ of an SPN of a trace net and a model net. The target transition is $t^*$, and a target place $p^*$ is a successor to $t^*$. For this system net, $M_{init,ext}=[p_{(\blacktriangleright,b)},p_0]$ and $M_{final,ext}=[p^*]$}
    \label{fig:extended-spn}
\end{figure}

\begin{definition}[Extended Synchronous Product Net]
    Let $SN_{\otimes}=((P_{\otimes},T_{\otimes}, \allowbreak F_{\otimes},\lambda_{\otimes}), \allowbreak M_{init,\otimes}, M_{final,\otimes})$ denote a synchronous product net of a trace net and an easy sound \footnote{A net is \textit{easy sound} if its complete finite prefix $\mathcal{P}$ contains at least one complete distributed run.} model net. An \textit{extended sychronous product net} of $SN_{\otimes}$ is another system net $SN_{\otimes}^{ext}=((P_{\otimes}^{ext},T_{\otimes}^{ext},F_{\otimes}^{ext},\lambda_{\otimes}^{ext}),  M_{init,\otimes}^{ext},M_{final,\otimes}^{ext})$ obtained by extending $SN_{\otimes}$ with a transition $t^*$ and a place $p^*$ such that $\pre{(t ^ *)}=M_{final,\otimes}$,  $\pre{(p ^ *)}=t ^ *$ and $\post{(p ^ *)}=\emptyset$. For such a system net, $M_{init,\otimes}^{ext} = M_{init,\otimes}$ and $M_{final,\otimes}^{ext} = [p ^ *]$.
\end{definition}

\Cref{fig:extended-spn} shows an extended SPN $SN_{\otimes}^{ext}$ of an SPN $SN_{\otimes}$ after adding the target transition and place depicted in dotted red. We shall be using this product net of an exemplary trace net (induced by a p-trace) and an easy sound model net shown in \Cref{fig:extended-spn}) in the remainder of this paper.

Assume a branching process of this extended product net. We define a \textit{target event} $e^*$ in its branching process that maps to the target transition, i.e., for which $h(e^*)=t^*$. The encoding trick ensures that $e^*$ is enabled only when the target marking corresponding to the final marking is reached. Therefore, its local distributed run represents a \textit{complete} distributed run leading to $e^*$. We refer to such a complete distributed run as an \textit{alignment run}.

\begin{definition}[Alignment Run]
\label{def:alignment-run}
    Let $SN_{\otimes}^{ext}$ denote an extended synchronous product net obtained by extending $SN_{\otimes}$ with a transition $t^*$ and a place $p^*$. An \textit{alignment run} $\upgamma=(B,E,F,h)$ is a complete distributed run of $SN_{\otimes}^{ext}$ leading to a target event $e^*$ such that the set $\{b \in B \mid \post{b} = \emptyset\}$ contains a single condition $b^*$ where $\pre{b^*}=e^*$ and $h(e^*)=t^*$. We denote the universe of all such alignment runs by $\mathcal{U}_{SN_{\otimes}^{ext}}$.
\end{definition}

% It can easily be proved that there exists at least one alignment run in an extended synchronous product net since there is always a complete distributed run in a trace net that does not touch any synchronous or model parts (this distributed run exists because a trace net is always easy sound \Cref{lem:trace-net-easy-soundness}) and the model net also contains a complete distributed run by definition. Therefore, an alignment run consisting of both the complete distributed runs in parallel until they enable a target event always exists. 

To ensure optimality, we use a \textit{scoring function over finite configurations}, a monotonic function assigning additive costs. This function allows comparing local configurations of target events, where a configuration with lower cost is preferred, effectively comparing distributed runs based upon costs of their member events. The function is used to define an \textit{optimal alignment run} in the following definition.

\begin{definition}[Scoring Function Over Finite Configurations]
    Let $\mathcal{C}_{all}$ denote the set of all finite configurations of a branching process $\upbeta=(B, E, F, h)$ of an extended SPN and \textit{cost} be a cost function on transitions of the SPN such that it assigns a constant positive cost to log and model transitions, whereas synchronous transitions have a cost of 0, and invisible transitions have a negligible positive cost. The \textit{scoring function over finite configurations} is defined as $s: \mathcal{C}_{all} \rightarrow \mathbb{N}_0$ such that $s(C) = \sum_{e \in C} cost(h(e)), C \in \mathcal{C}_{all}$. A scoring function is \textit{monotonic} since $\forall C,C' \in \mathcal{C}_{all}, C \subseteq C' \implies s(C) \leq s(C')$.
\end{definition}

% We extend the the scoring function $s$ to configurations $C \oplus \{e^*\}, C \in C_{all}$ by setting $s(C \oplus \{e^*\}) = s(C)$. Think of it as considering the cost of the target event as 0.
% Using this cost function, an \textit{optimal alignment run} can be defined as follows:

\begin{definition}[Optimal Alignment Run]
     Let $SN_{\otimes}^{ext}$ denote an extended SPN and let $\upbeta$ denote its branching process. Let $s: \mathcal{C}_{all} \rightarrow \mathbb{N}_0$ be a scoring function over the finite configurations of $\upbeta$. An \textit{optimal alignment run} $\upgamma^{opt} \in \mathcal{U}_{SN_{\otimes}^{ext}}$ is an alignment run of $SN_{\otimes}^{ext}$ leading to a target event $(e^*)^{opt}$ such that for all the other alignment runs $\upgamma \in \mathcal{U}_{SN_{\otimes}^{ext}}$ leading to target events $e^*_1,...,e^*_n$ respectively, it holds that $s([(e^*)^{opt}]) \leq s([e^*_i]), 1 \leq i \leq n$.
\end{definition}

To obtain an optimal alignment run, we unfold $SN_{\otimes}^{ext}$ (using the $ERV$ unfolding algorithm described in \Cref{sec:erv-unfolding-algo}) until a target event is pulled out from the queue of events. At this point, the unfolding process can be terminated and a local distributed run associated with $e^*$ is retrieved. This local distributed run is an alignment run as defined in \Cref{def:alignment-run}. Its alignment cost is given by $s([e^*])$.

Note that in the unfolding algorithm, an adequate order is used to order events in a queue. Using different adequate orders leads to different results, i.e., different complete distributed runs. Here we choose a \textit{cost-based adequate order} that compares finite configurations based upon their costs.

\begin{definition}[Cost-based Partial Order $\triangleleft_{c}$]
\label{def:cost-based-po}
    Let $C,C'$ be two finite configurations of a branching process $\upbeta$ of an extended SPN and $s$ be the scoring function. Let $\gg$ be a total order on the events of $\upbeta$ such that for any two events $e$ and $e'$, $e \gg e'$ if $e$ was added earlier to the prefix than $e'$. Given a set $E$ of events, $\phi(E)$ denotes the sequence of events that is ordered according to $\gg$. It is easy to see that $\gg$ is a well-founded relation. A \textit{cost-based partial order} $\triangleleft_c$ is defined such that $C \triangleleft_c C'$ if and only if:
    \begin{itemize}
        \item $s(C) < s(C')$, or
        \item $s(C)=s(C')$ and $|C| < |C'|$, or 
        \item $s(C)=s(C')$ and $|C| = |C'|$ and $\phi(C) \gg \phi(C')$ \footnote{We say $\phi(E_1) \gg \phi(E_2)$ if $\phi(E_1)$ is lexicographically smaller than $\phi(E_2)$ with respect to the total order $\gg$}.
    \end{itemize}
\end{definition}

That $\triangleleft_c$ is irreflexive, asymmetric and transitive is obvious from the definition. Hence, $\triangleleft_c$ is also a strict partial order. In addition to ordering the priority queue, the chosen adequate order is used to identify cut-off events. For correctness proofs to work, i.e., for the $ERV$ algorithm to correctly identify cut-off events, a cost-based order $\triangleleft_c$ defined as above must be \textit{adequate} \cite{esparza1996improvement}. It can be proven that $\triangleleft_c$ satisfies the three properties of adequate orders (cf. \Cref{adequacy} in Appendix) and hence can be used to correctly unfold extended SPNs with respect to the target marking.

Next, we present the $ERV[\triangleleft_c]$ (along with $ERV[\triangleleft_h]$ as discussed later in \Cref{sec:directed-unf}) \textit{unfolding algorithm}, an on-the-fly variant of $ERV$ algorithm \cite{esparza1996improvement} to produce optimal alignment runs. It is implemented as the \textit{UnfoldSyncNet} procedure in Algorithm \ref{alg:ervc-unfolding}. It differs from the standard $ERV$ unfolding algorithm in that it terminates when a target event is pulled out from the queue (based upon the chosen $stopAtFirst$ flag), returning an optimal alignment run as an output. Additionally, it uses $\triangleleft_c$ as its adequate order.

\subsection{$ERV[\triangleleft_c]$ Unfolding Algorithm}
\label{sec:erv-unfolding-algo}

\SetKwComment{Comment}{/* }{ */}
\SetKw{KwRet}{return}
\begin{algorithm}[t]
\scriptsize

    \caption{UnfoldSyncNet - $ERV[\triangleleft_c]$ (and $ERV[\triangleleft_h]$, cf. \Cref{sec:directed-unf}) }
    \label{alg:ervc-unfolding}
    \KwIn{$SN_{\otimes}^{ext}=((P_{\otimes}^{ext}, T_{\otimes}^{ext}, F_{\otimes}^{ext}, \lambda_{\otimes}^{ext}), M_{init,\otimes}^{ext}, M_{final,\otimes}^{ext})$, $stopAtFirst \in \mathbb{B}$}
    \KwOut{$alignmentRuns \subseteq \mathcal{U}_{SN_{\otimes}^{ext}}$, $lowestCost \in \mathbb{N}_0$}
    \textit{cut-off, pe, finalEvents, $\beta$} $\leftarrow \emptyset$ {\color{gray}\tcp*{Initialize the set of cutoff events, priority queue ordered by $\triangleleft_c$ (or $\triangleleft_h$ for $ERV[\triangleleft_h]$), the set of target events with minimal $s([e])$, and the prefix $\beta = (B,E,F,h)$}}
    $imarks \leftarrow \{M_{init,ext}: \bot\}${\color{gray}\tcp*[r]{Lookup table of induced markings initialized with the initial marking induced by a dummy event $\bot$}}
    \For({\color{gray}\tcp*[f]{Add initial conditions to prefix}\color{black}}){$p \in M_{init,ext}$}{
        $B \leftarrow B \ \cup$ \{a new condition mapped to place $p$\}
    }
    \color{blue}calculate all possible extensions of $B$\\
    \color{black}\While {$pe \neq \emptyset$ }{
        $e \leftarrow Min\{pe\}$ {\color{gray}\tcp*{Retrieve an event $e$ s.t. [$e$] is minimal w.r.t. $\triangleleft_c$ (or $\triangleleft_h$ for the $ERV[\triangleleft_h]$ variant)}}
        \uIf(\color{gray}\tcp*[f]{A target event found}\color{black}){$h(e)=t^*$} { 
        $alignmentRuns \leftarrow alignmentRuns \cup \{\mathcal{O}_e\}$ {\color{gray}\tcp*{A local distributed run $\mathcal{O}_e$ of $e$ (cf. \Cref{def:local-dist-run}) is added}}
            $lowestCost \leftarrow s([e])$\\
            \uIf{$stopAtFirst = true$}{
                \textbf{break} 
            }
            \lElse{
                \textbf{continue}
            }
        }
    
        \uIf{$[e] \cap \text{cut-off} = \emptyset$}{
            \For(\color{gray}\tcp*[f]{Add $e$'s postset conditions one by one}\color{black}){$p \in \post{h(e)}$}{
                $c \leftarrow$ a new condition mapped to $p$
                extend from event $e$ to condition $c$ in the prefix $\beta$
            }
            \color{black}
            \uIf(\color{gray}\tcp*[f]{check if $Mark([e])$ is already present in $imarks$}\color{black}){$e$ is a cut-off event w.r.t. $\triangleleft_c$ (or $\triangleleft_h$ for $ERV[\triangleleft_h]$)}{
                $\textit{cut-off} \leftarrow \textit{cut-off} \ \cup \{e\}$
            }
            \uElse{
                \color{blue}calculate all possible extensions of $B$\\
                \color{black}add $Mark([e])$ to $imarks$
            }
        }
    }
   \KwRet{$alignmentRuns, lowestCost$}
\end{algorithm}

Given an extended SPN $SN_{\otimes}^{ext}=((P_{\otimes}^{ext}, T_{\otimes}^{ext}, F_{\otimes}^{ext}, \lambda_{\otimes}^{ext}), \allowbreak M_{init,\otimes}^{ext}, M_{final,\otimes}^{ext})$ and a boolean flag for $stopAtFirst$ to indicate if only one optimal alignment run is to be computed or all, we can obtain (all) optimal alignment run(s) (and thus, all optimal alignment orders) from the initial marking $M_{init,\otimes}^{ext}$ to the final marking $M_{final,\otimes}^{ext}$ using Algorithm \ref{alg:ervc-unfolding}.

Similar to the classic $ERV$ unfolding algorithm, $ERV[\triangleleft_c]$ builds a prefix of events and conditions using a priority queue ordered by an adequate order ($\triangleleft_c$). Unlike the classic version, it adds steps (lines 8-13) to handle target events enabling the target marking. When such an event is retrieved, its local distributed run is stored in \textit{alignmentRuns} (line 9). If only one alignment run is needed (line 11), the loop ends early, returning the result. Otherwise, the algorithm checks if the event's local configuration contains a cut-off event (line 14) using a lookup table $imarks$ for easy O(1) retrieval. If no cut-off exists, the prefix is extended with the event's postset conditions (Lines 15-16). Additionally, if $e$ itself is a cutoff-event, it is added to the set of cutoff events (Lines 17-18). Any new possible extensions are calculated (line 20), added to the prefix, and $imarks$ is updated (line 21). The process repeats until the queue is empty. 

The part in blue in the algorithm, i.e. the calculation of all possible extensions, is the one crucial to implementation. It deals with the problem of finding events that are enabled by the entire set of input conditions while ensuring that they are not in causal or conflict relation with each other. Those identified events are thus added to the prefix and priority queue. 
% With a brute force approach where \textit{all} subsets of conditions are considered for extension, there are $2^{|B|-1}$ ones for each current prefix size of $|B|$ to consider. Of course, a larger prefix size quickly leads to a combinatorial explosion. 
Efficient conflict/causal detection while downsizing the number of combinations to check for is key to managing the combinatorial explosion inherent in this part of the algorithm. Roemer in his work \cite{romer2000theorie} already proposed ways to optimize this part using intelligent data structures and procedures. In our work, we implement the same techniques.

\subsection{Directing the Unfolding}

\label{sec:directed-unf}

In the context of $ERV[\triangleleft_c]$ algorithm, we retrieve events from the queue based upon the total cost of mapped transitions in its local configuration, i.e., $s([e])$; equivalent to selecting events based upon the transitions fired so far in its local distributed run. For efficiency, we also consider an \textit{underestimate} of the cost of transitions remaining till the target event is reached. Thus, when selecting events from the queue, we additionally favor those "closer" to $t^*$, resulting in a \textit{directed unfolding} strategy. To this end, we use a \textit{heuristic function} on configurations to guide the unfolding process towards a target configuration. Its value is an estimate of the distance (where the notion of distance is related to the total cost of transitions fired) between configurations. Thus, in the context of $ERV[\triangleleft_c]$ algorithm, orderings can be constructed upon the values of a function $f: \mathcal{C}_{all} \rightarrow \mathbb{N}_0$ composed of two parts $f(C) = s([e]) + est(C)$, where $s([e])$ is the total cost of mapped transitions of $[e]$ and $est(C)$, the heuristic value, estimates the distance from $Mark(C)$ to the target marking $[p^*]$. We use a heuristic function that exploits the marking equation \cite{carmona2018conformance} to underestimate the remaining cost to $[p^*]$.

\begin{definition}[Heuristics-based Partial Order $\triangleleft_h$]
    Let $C,C'$ be two finite configurations of a branching process $\upbeta$ of an SPN. Let $f(C) = s([e]) + est(C)$ be a function on finite configurations where $s$ is a scoring function and $est$ is a heuristic function. Let $\triangleleft_c$ be a cost-based adequate order over finite configurations. A \textit{heuristics-based partial order} $\triangleleft_h$ is defined such that:
    \begin{align*}
    C \triangleleft_h C' \text{  iff  } \begin{cases}
    			f(C) < f(C'), & \text{if $f(C) \neq f(C')$}\\
                    C \triangleleft_c C', & \text{otherwise.}
    		 \end{cases}
    \end{align*}
\end{definition}

% This results in a \textit{breadth-first} search sorting criterion on configurations with an equal value of $f$. 
Given that the first part of $f$ (in our case, $s$) is a monotonic function and $est$ is an admissible heuristic function, Bonet et al. \cite{bonet2008directed} proved that a partial order defined as above belongs to a family of \textit{semi-adequate} orders, which still enables us to use the $ERV[\triangleleft_c]$ algorithm with the same definition of cut-off events. Here, we simply replace $\triangleleft_c$ in $ERV[\triangleleft_c]$ (\Cref{alg:ervc-unfolding}) by $\triangleleft_h$ to obtain another algorithm variant, referred to as $ERV[\triangleleft_h]$ \textit{algorithm}. The only changes in $ERV[\triangleleft_h]$ occur in Lines 7 and 17 where an adequate (semi-adequate, in this case) order is used.  

\subsection{Building and Visualizing Optimal Alignments from Optimal Alignment Runs}

\begin{figure}[tb]
    \centering
    \includegraphics[width=.9\linewidth]{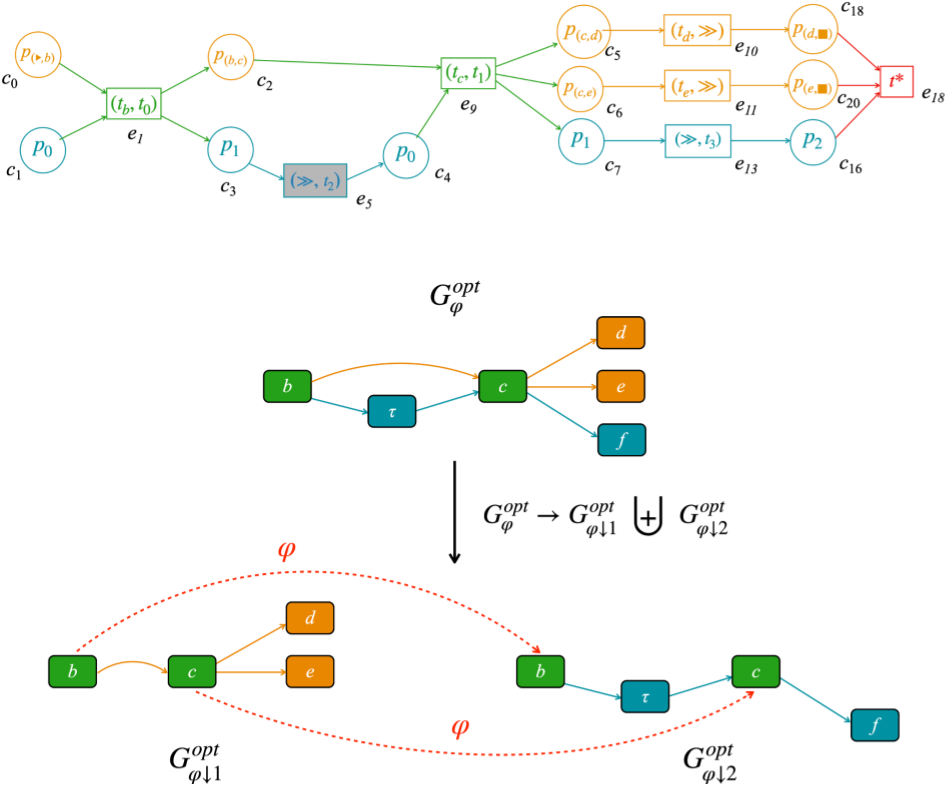}
    \caption{Optimal alignment order (middle), derived from the alignment run (top) of the extended SPN $SN_{\otimes}^{ext}$, with partial order decomposition showing the u-alignment (bottom). $G^{opt}_{\upvarphi \downarrow 1}$ is isomorphic to the p-trace and $G^{opt}_{\upvarphi \downarrow 2}$ is the partial order from the model net run (cf. \Cref{fig:extended-spn}). Dotted red arrows show the mapping of the alignment function.}
    \label{fig:decomposition}
\end{figure}

Building a u-alignment from an optimal alignment run, referred to as the \textit{optimal u-alignment}, is a two-step process:

\begin{enumerate}
    \item Construction of an optimal alignment order by using the events and conditions of the net corresponding to an optimal alignment run.
    \item Decomposition of the optimal alignment order to obtain two partial orders, one corresponding to the p-trace and the other to a model run, while setting our alignment function at the same time. This gives us an optimal u-alignment, our final artefact. 
\end{enumerate}

% In the first step, assume we have an optimal alignment run $\upgamma^{opt} = (B^{opt}, E^{opt}, \allowbreak F^{opt}, h^{opt})$ of an extended synchronous product net $SN_{\otimes}^{ext}=((P_{\otimes}^{ext},T_{\otimes}^{ext},W_{\otimes}^{ext}, \allowbreak \lambda_{\otimes}^{ext}),  M_{init,\otimes}^{ext}, M_{final,\otimes}^{ext})$ where $e^* \in E^{opt}$ is a target event. An optimal alignment order $G_\upvarphi^{opt} = ((M, \prec_\upvarphi), l_\upvarphi)$ is a partial order with:

% \begin{itemize}
%     \item $M=\{h^{opt}(e) \mid e \in E^{opt} \textbackslash \{e^*\} \}$,
%     \item $\prec_\upvarphi = \{ (h^{opt}(e_1),h^{opt}(e_2)) \mid e_1,e_2 \in E^{opt} \textbackslash \{e^*\} \text{ such that } \exists b \in B^{opt} \text{ for which } \allowbreak (e_1,b) \in F^{opt} \wedge (b,e_2) \in F^{opt}\}$,
%     \item $l_{\upvarphi}: M \rightarrow \mathcal{L}^\uptau$ where $l_\upvarphi(m) = \lambda_{\otimes}^{ext}(m), m \in M$
% \end{itemize}

In the first step, to build an optimal alignment order $G_{\upvarphi}^{opt}$ from an optimal alignment run $\upgamma^{opt}$, remove the target event $e^*$, map each event to a labeled node representing an alignment move, and create dependencies based on conditions, classifying them as log or model dependencies. In the second step, we decompose an alignment order into two partial orders while setting the alignment function. \Cref{fig:decomposition} shows the mechanism through an illustration. On top we see an optimal alignment run through the example extended SPN from \Cref{fig:extended-spn} using $ERV[\triangleleft_c]$ or $ERV[\triangleleft_h]$. Below it, is the transformed optimal alignment order $G^{opt}_\upvarphi$. $G^{opt}_\upvarphi$ is decomposed into different partial orders $G^{opt}_{\upvarphi \downarrow 1}$ and $G^{opt}_{\upvarphi \downarrow 2}$, i.e., $G^{opt}_{\upvarphi} \rightarrow G^{opt}_{\upvarphi \downarrow 1} \biguplus G^{opt}_{\upvarphi \downarrow 2}$. The alignment function $\upvarphi$ relates synchronous nodes `b' and `c' in the two partial orders. Moreover, notice that $G^{opt}_{\upvarphi \downarrow 1}$ is isomorphic to the p-trace we computed our alignment for (observe the p-trace in \Cref{fig:extended-spn}). Thus, our final u-alignment is $(G^{opt}_{\upvarphi \downarrow 1}, G^{opt}_{\upvarphi \downarrow 2}, \upvarphi)$.

% One way to interpret this representation is to notice that the decomposition highlights events which occur synchronously in the log and model while preserving the order between events in both. Therefore, it allows for a comparison of both, highlighting where the log and model align in event order and concurrency.

% \begin{figure}
% \centering
%      \includegraphics[width=0.7\textwidth]{figures/cviews-running-eg.png}
%      \caption{Chevron-based concurrency views of $G^{opt}_{\upvarphi \downarrow 1}$ and $G^{opt}_{\upvarphi \downarrow 2}$. Mapping by the alignment function (shown in dashed red) is still preserved} 
%      \label{fig:cviews-running-eg}
% \end{figure}

% For our example from the section, this is shown in \Cref{fig:cviews-running-eg}. Notice that the alignment function still preserves the mapping between synchronous moves `b' and `c', shown in red and blue chevrons, respectively.

The decomposed partial orders $G^{opt}_{\upvarphi \downarrow 1}$ and $G^{opt}_{\upvarphi \downarrow 2}$ are visualized into chevron-based views as exemplified in \Cref{fig:alignment-function}. This visualization method effectively highlights alignment between log and model events but falls short in identifying \textit{misalignments}, particularly regarding dependencies. When comparing a model with a log variant, we aim to identify events present in reality but missing in the model (and vice versa) and to detect violations in event orders. These observations allow us to classify misalignments into two main categories. In the following, we assume a u-alignment $(G^{opt}_{\upvarphi \downarrow 1}, G^{opt}_{\upvarphi \downarrow 2}, \upvarphi)$ where $\prec^{opt-}_{\upvarphi \downarrow 1}$, $\prec^{opt-}_{\upvarphi \downarrow 2}$ are the edge relations of $G^{opt-}_{\upvarphi \downarrow 1}$ and $G^{opt-}_{\upvarphi \downarrow 2}$, respectively, where $G^{-}$ denotes the transitive reduction of a DAG $G$.

\begin{enumerate}
    \item \textbf{Missing events and dependencies}: Events and dependencies in the log but not in the model. Missing dependencies are relations \( (m_1, m_2) \in \prec^{opt-}_{\upvarphi \downarrow 1} \) with \( (m_1, m_2) \notin \prec^{opt-}_{\upvarphi \downarrow 2} \), representing log dependencies without corresponding model dependencies. Missing events are log moves of the form  $(m_1, \gg)$.

    \item \textbf{Undesired events and dependencies}: Events and dependencies in the log but not in the model. Missing dependencies are relations \( (m_1, m_2) \in \prec^{opt-}_{\upvarphi \downarrow 2} \) with \( (m_1, m_2) \notin \prec^{opt-}_{\upvarphi \downarrow 1} \), representing log dependencies without corresponding model dependencies. Missing events are log moves of the form \( (\gg, m_2) \).
\end{enumerate}

In $G^{opt}_{\upvarphi \downarrow 1}$ and $G^{opt}_{\upvarphi \downarrow 2}$ from \Cref{fig:decomposition}, there are in total three missing dependencies corresponding to all dependencies visible in $G^{opt}_{\upvarphi \downarrow 1}$, three undesired dependencies corresponding to all dependencies in $G^{opt}_{\upvarphi \downarrow 2}$. Moreover, there are two missing events corresponding to `d' and `e' (orange nodes) and two undesired events corresponding to `$\uptau$' and `f' (blue nodes).

\section{Tool Support}

The $ERV[\triangleleft_c]$ and ($ERV[\triangleleft_h]$) algorithms have been implemented\footnote{\url{https://github.com/ariba-work/cortado}} in \textit{Cortado} \cite{schuster2021cortado}, a tool for interactive process discovery. %, which builds on the Python process mining library \textit{PM4Py} \cite{berti2023pm4py}. 
Cortado traditionally checks conformance by computing all possible sequentializations of each partially ordered trace variant and aligning each sequence using the classic sequential alignment method \cite{van2012replaying}. The final alignments for each variant are aggregates of the sub-alignments. Our implementation replaces this sequentialization-based approach with unfolding-based alignments, adding new features.

\begin{figure}[t]
    \centering
    \includegraphics[width=.75\linewidth]{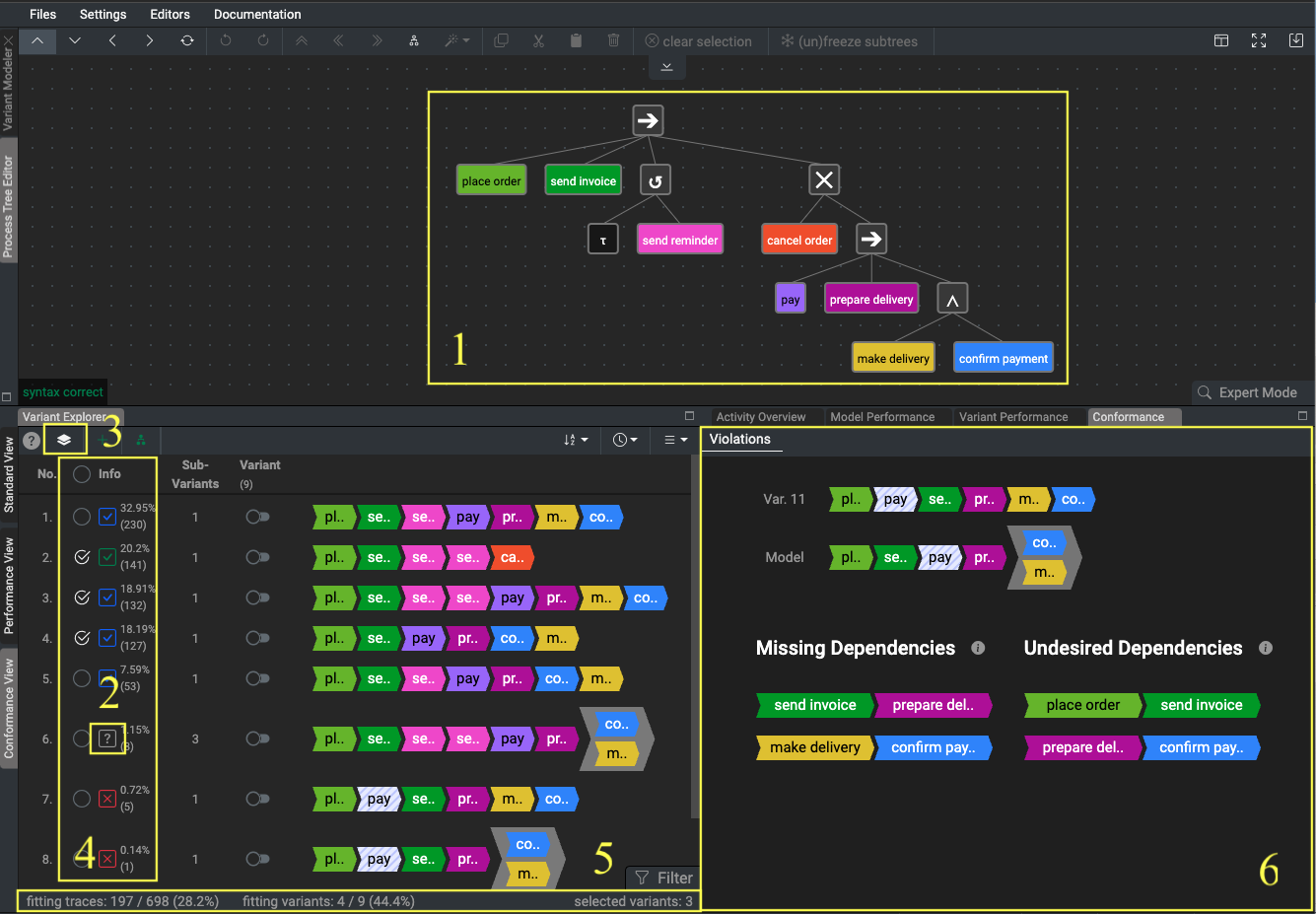}
    \caption{Overview of the Conformance Analysis interface in Cortado}
    \label{fig:interface-cortado}
\end{figure}

 We present a visual overview of the modified Conformance Analysis interface in \Cref{fig:interface-cortado}. Once an initial model based on selected variants is available, users can compute unfolding-based alignments for new variants. The model, shown in the process tree editor (1), allows alignment computation by clicking one of the two buttons to (re)calculate conformance statistics (2), (3). After computation, one of three statuses appears (4): 1. variant fits the model, 2. variant does not fit, or 3. variant fits but with order deviations (our contribution). Conformance statistics is shown at the bottom (5). For a more detailed analysis, the Violations tab (6) in `Conformance View' displays chevron-based concurrency visualizations of each selected variant and the lowest cost model run corresponding to that variant. Missing events are highlighted with stripes, and synchronous events can be paired by hovering over them. Missing and undesired dependencies are listed as pairs of chevrons, showing where dependencies should exist but don't in the model (e.g., `send invoice' followed by `prepare delivery') or vice versa. 
 
 % Dependencies related to missing events are filtered by default, and violations from skipped events in the model are excluded.

\section{Evaluation}

We evaluate $ERV[\triangleleft_c]$ and $ERV[\triangleleft_h]$ against the classic approach to partial alignments based on $A^*$ algorithm. Since we are particularly interested in the performance of our algorithm in settings of high parallelism, we split our experiments into two: 1). where we generate data artificially in order to control the degree of parallelism and compare different algorithms in the respective settings (\Cref{sec:artificial-eval}), and 2). where we evaluate using real-life event logs (\Cref{sec:real-life-eval}). In each part, we present our experimental setup and discuss the results. Both $ERV[\triangleleft_c]$ and $ERV[\triangleleft_h]$ are implemented in a forked repository\footnote{\url{https://github.com/ariba-work/cortado}} of the open-source software tool for process mining, \textit{Cortado} \cite{schuster2021cortado}. Furthermore, to compare our results with the classic approach to partial alignments (referred to as $Classic \ PA$ from hereon) \cite{lu2015conformance}, we implement their algorithm in a separate Github repository\footnote{\url{https://github.com/ariba-work/classic-pa}}.

\subsection{Using Artificial Event Logs}
\label{sec:artificial-eval}

To examine the impact of parallelism on computation time, we conducted an experiment to test our hypothesis: as concurrency in models increases, the runtime overhead for exploring interleavings in $Classic \ PA$'s reachability graph grows larger than the overhead for calculating extensions in $ERV[\triangleleft_c]$ and $ERV[\triangleleft_h]$. Consequently, our proposed algorithms are expected to outperform $Classic \ PA$ at higher parallelism levels.

\subsubsection{Experimental Setup}

The experimental workflow involves generating 8 process trees with varying parallelism levels (0 \% to 70\%) using \textit{PTAndLogGenerator} \cite{jouck2019generating} such that all trees depict block-structured Petri nets without loops and/or duplicate labels. We then convert the trees into Petri nets, and reduce silent transitions, using different ProM\footnote{\url{https://promtools.org/}} plugins. Event logs with 500 traces are simulated for each net using the plugin \textit{StochasticPetriNets}. Noise in levels of 0 \% to 50 \% is inserted using an existing noise-insertion technique proposed in \cite{van2021naturalnoise}. Finally, conformance checking is performed using $ERV[\triangleleft_c]$, $ERV[\triangleleft_h]$, and $A^*$-based $Classic \ PA$ on each of the simulated event log. 
% The scripts, written in Python, for inserting noise and triggering the conformance checking experiments are available in the same forked Github repository of Cortado.

\subsubsection{Results}

\begin{figure}
    \centering
    \includegraphics[width=.75\textwidth]{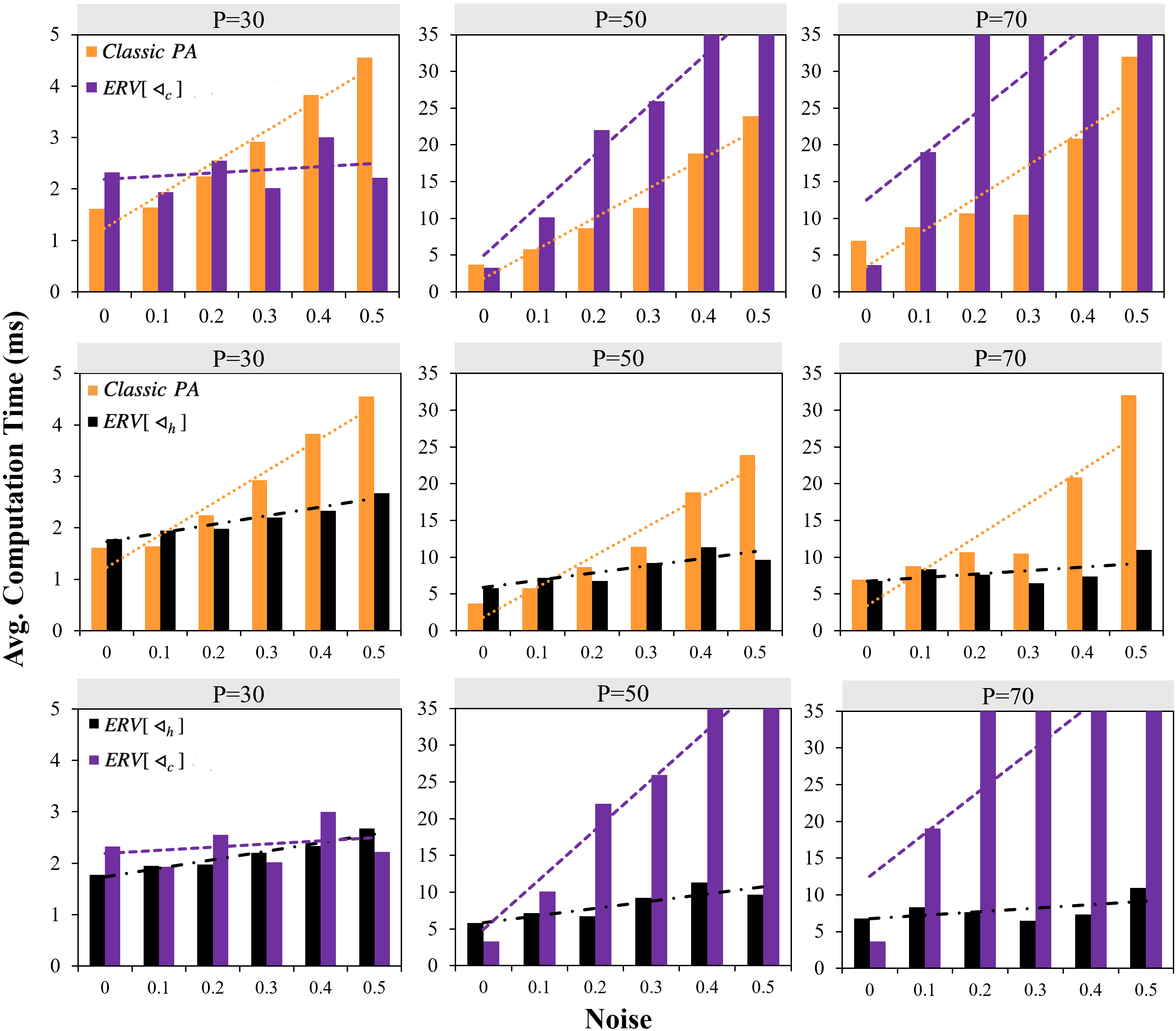}
    \caption{Average computation time (ms) per percentage of noise, plotted per level of parallelism (30\%, 50\% and 70\%). The level of parallelism is shown on top of the plots. In each row, the comparison is shown for two algorithms at a time.}
    \label{fig:parallelism-vs-time}
\end{figure}

In \Cref{fig:parallelism-vs-time}, we compare $ERV[\triangleleft_c]$, $ERV[\triangleleft_h]$, and $Classic \ PA$ for varying noise and parallelism levels, focusing on average computation times. Results are shown for parallelism levels of 30\%, 50\%, and 70\%, where significant differences were observed. The x-axis represents noise percentage in event logs, and the y-axis shows computation times in milliseconds, with linear regression lines illustrating trends. At 70\% parallelism, $ERV[\triangleleft_h]$ outperforms other variants, showing robustness. On the other hand, $ERV[\triangleleft_h]$ starts to slow down even at 50\%, due to larger prefixes generated, caused by increased events in high-parallelism models and the lack of heuristics to limit prefix expansion. At 30\% parallelism, all variants perform almost similarly, with $ERV$ variants excelling under higher noise levels. These findings support the hypothesis that unfolding-based alignments are more robust as parallelism increases.

\subsection{Using Real Life Event Logs}
\label{sec:real-life-eval}

\subsubsection{Experimental Setup}

For this experiment, we selected the BPI Challenge 2012 (BPIC 12) \cite{bpic12} dataset, which contains partial event data suitable for validating our approach. The BPIC 12 dataset contains 262,200 events across 13,087 cases and 4,366 variants. It includes both start- and end-timestamps, with case lengths ranging from 3 to 175 events.

\begin{table}[t]
\caption{Statistics of the models discovered by Infrequent Inductive Miner. Column names are abbreviated (\#Pl: number of places, \#Tr: number of transitions, \#Si: number of silent transitions, ECyM: Extended Cyclomatic Metric)}
\label{tab:model-metrics}
  \centering
  \scriptsize
  \setlength{\tabcolsep}{18pt}
  \begin{tabular}{lccccc}
    \toprule
    \textbf{BPIC 12 Model} & \textbf{\#Pl} & \textbf{\#Tr} & \textbf{\#Si} & \textbf{ECyM} \\
    \midrule
    \textbf{noise thr. 0.05} & 37 & 50 & 33 & 1899 \\
    \textbf{noise thr. 0.5} & 26 & 27 & 11 & 289 \\
    % \textbf{noise thr. 0.8} & 19 & 20 & 3 & 83 \\
    \bottomrule
  \end{tabular}
\end{table}

For conformance checking experiments, process models were discovered using the \textit{Infrequent Inductive Miner} algorithm in ProM with noise thresholds of 5\% and 50\% where higher thresholds simplify models by filtering out rare traces. \Cref{tab:model-metrics} shows the number of places, transitions, silent transitions and additionally the \textit{Extended Cyclomatic Metric} (ECyM) \cite{LassenA09} of the mined models. This metric is calculated using the reachability graph of Petri nets where higher values indicate more reachability paths.

\subsubsection{Results}

\begin{figure}[tb]
    \centering
    \begin{subfigure}[]{.49\linewidth}
         \centering
         \includegraphics[width=\linewidth]{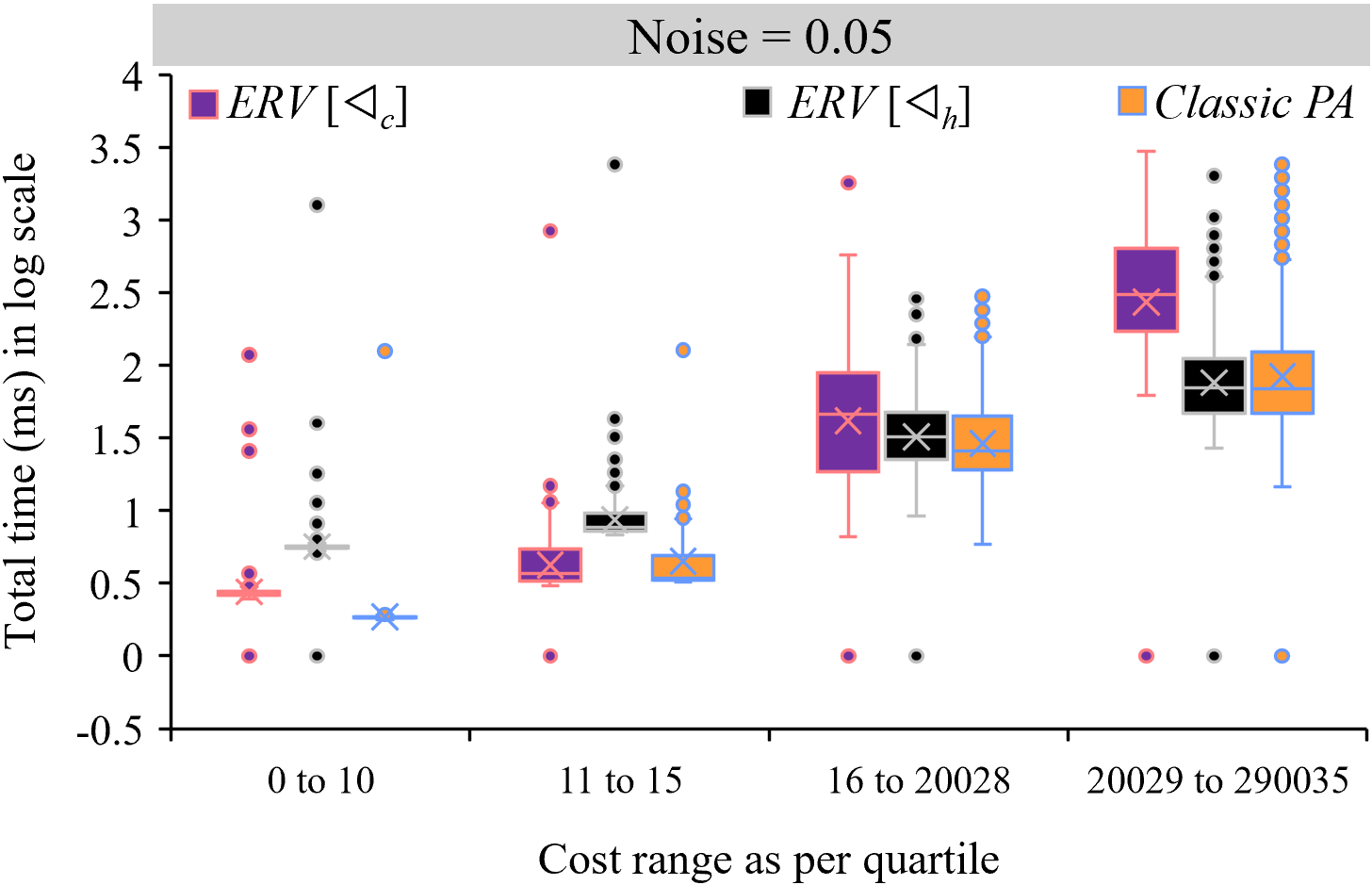}
         \label{fig:cost-vs-time-bpi12-0.05}
     \end{subfigure}
     \begin{subfigure}[]{.49\linewidth}
         \centering
         \includegraphics[width=\linewidth]{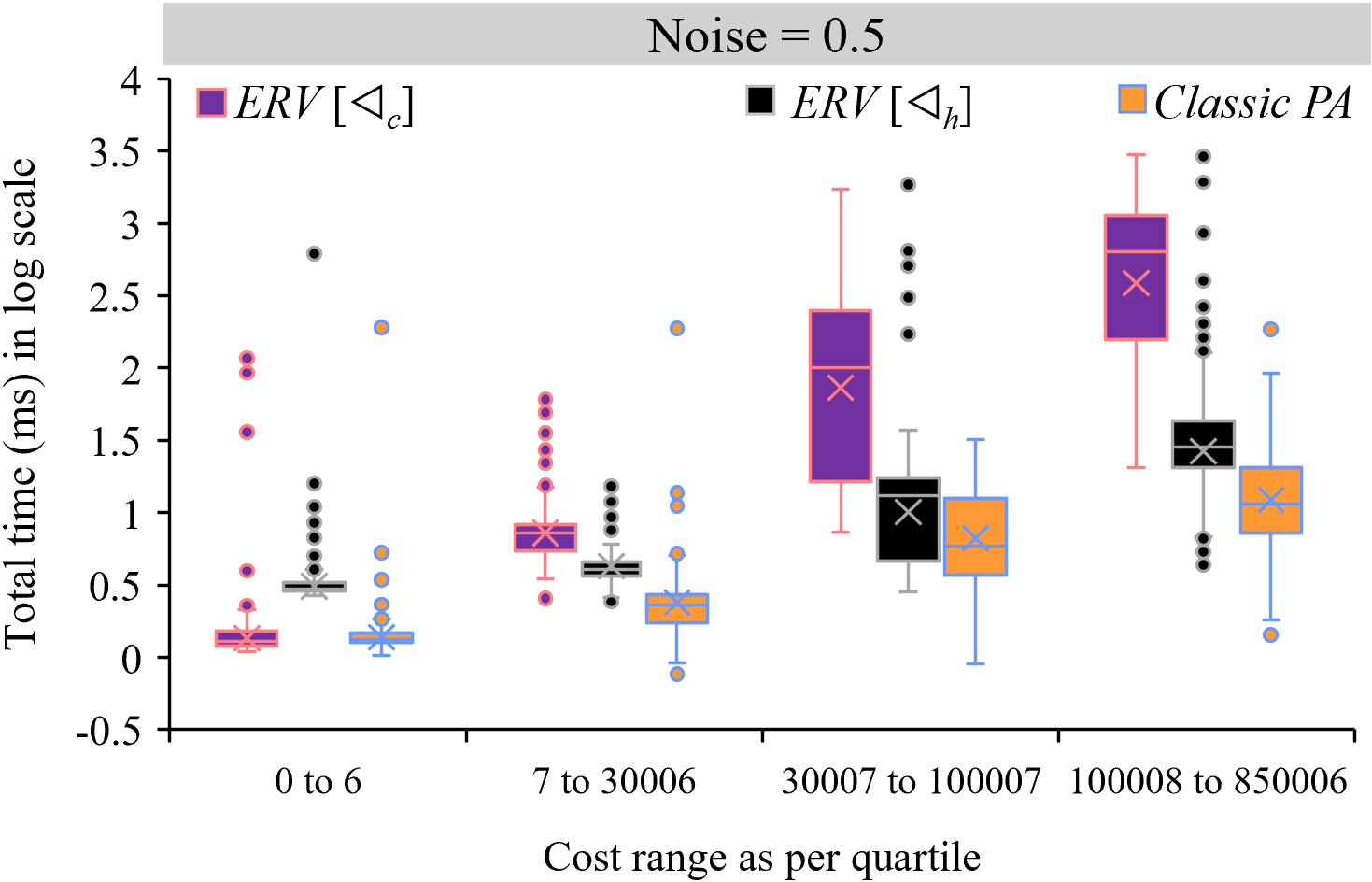}
         \label{fig:cost-vs-time-bpi12-0.5}
     \end{subfigure}
     \caption{Runtime comparison regarding deviation costs for different algorithms applied on models with different noise thresholds for BPIC 12 event log. Timeout for individual alignments for single traces is set to 3 seconds. $ERV[\triangleleft_c]$ generally performs the worst, while $ERV[\triangleleft_h]$ is more robust at the lowest noise threshold.}
    \label{fig:cost-vs-time-bpi12}
\end{figure}

\begin{figure}[tb]
    \centering
    \includegraphics[width=.8\linewidth]{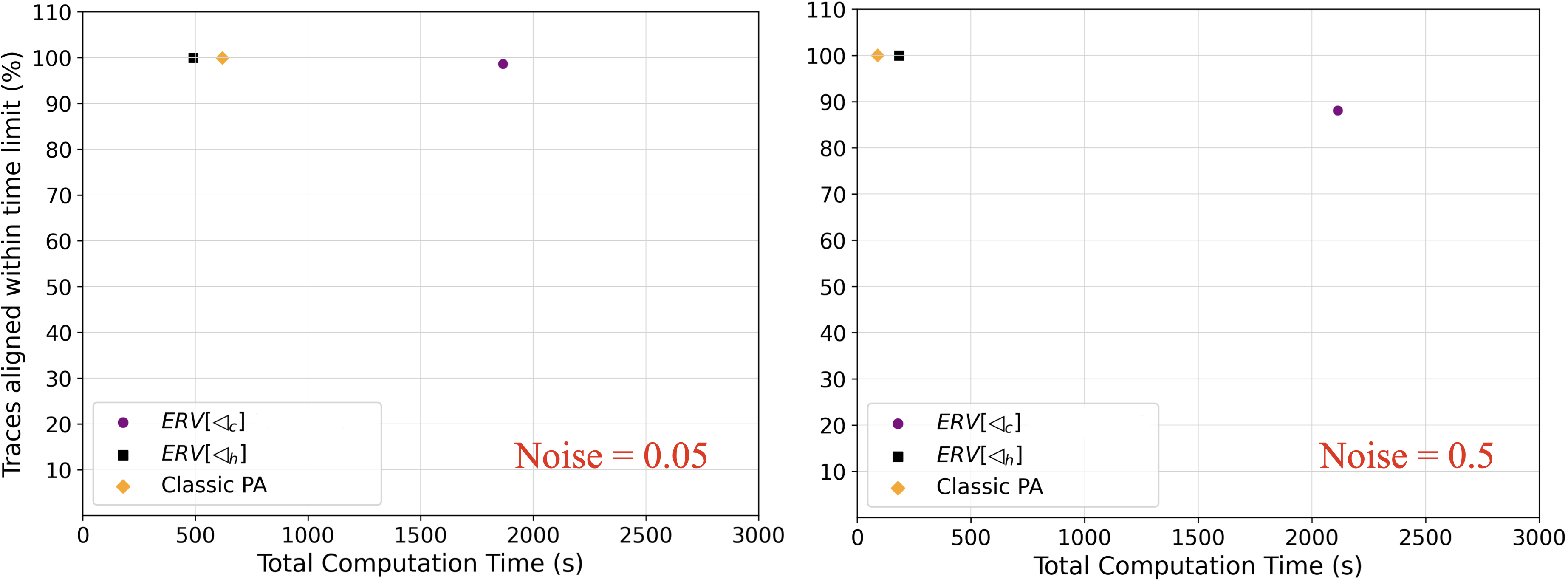}
    \caption{Percentage of traces aligned within time limit by different algorithms, along with the total computation time. The time limit applies to individual traces, and total time is for aligning the entire BPIC 12 event log with its model at various noise thresholds.}
    \label{fig:percent-traces-aligned}
\end{figure}

In \Cref{fig:cost-vs-time-bpi12}, we compare the runtime of $ERV[\triangleleft_c]$, $ERV[\triangleleft_h]$, and \textit{Classic PA} across cost ranges and noise thresholds. Box plots show that runtime increases approximately exponentially with deviation costs, with lower noise threshold models taking longer. In both plots, it is apparent that $Classic \ PA$ performs better for lower deviation costs. At 0.5 noise threshold, \textit{Classic PA} still outperforms both $ERV$ variants at higher deviation costs. This can be attributed to the higher number of non-synchronous transitions in the SPN of a higher noise threshold model, which necessitates exploring more equal-cost events. This results in broader prefixes and increased computation times. At 0.05 noise, $Classic \ PA$ performs better for low deviation costs (0-10), while $ERV[\triangleleft_h]$ is slower but excels for higher costs. The difference between $ERV[\triangleleft_h]$ and  $Classic \ PA$ is not \textit{so} apparent in this plot since runtime in the plots discussed so far does not take into account the traces for which alignment computations timed out (> 3 seconds). However, the total computation for this model differs by nearly 130 seconds. As shown in \Cref{fig:percent-traces-aligned}, $ERV[\triangleleft_h]$ and $Classic \ PA$ align nearly all traces across thresholds, while $ERV[\triangleleft_c]$ frequently times out at higher thresholds.

% \subsection{Threats to validity}

% In our exploratory experiment to gather evidence for the argument that $ERV[\triangleleft_c]$ algorithms are faster at higher parallelism, we only focused on block-structured Petri net models. This was done in order to minimize the effects other parameters may have. By allowing loops and duplicate visible transitions, a wider range of parameters can be examined to measure the performance of the respective algorithms. We also select the number of transitions in artificially created models from a fixed triangular distribution. The effect other distributions may have on performance is not clear. In our experiments on real-life event logs, we conducted our experiments using only one event log containing partially ordered event data. An even more extensive set of event logs can be used for further analysis. Particularly, an event log with higher concurrency patterns can be considered. We employed a unit-cost function in all our experiments. The computation times and results can vary significantly based on the selected cost function. It would be beneficial to explore other cost functions. This exploration should also take into account use cases where informed cost functions can be selected based on process knowledge.

\section{Conclusion}

This paper presented Petri net unfolding as an efficient solution for computing alignments from partially-ordered event data, overcoming the state space explosion of traditional reachability graph-based methods. Our approach simplifies alignment computation into a single step, avoiding the need for reconstructing alignments from totally-ordered sequences. We also introduce a graph-based visualization for unfolding-based alignments, enhancing the interpretability of log-model relationships. Experimental results show that the computation of unfolding-based alignments outperforms traditional alignment methods in high-parallelism and complex models. While there is some computational overhead due to the calculation of possible extensions in low-parallelism settings, the runtime and robustness of our proposed algorithms in handling concurrency validate its effectiveness. Moreover, with our visualization technique, it is possible to obtain diagnostics over missing and undesired events and dependencies on partially-ordered alignments that have not been possible otherwise. Implemented in \textit{Cortado}, this work provides both theoretical and practical contributions to conformance analysis. Future work will focus on optimizing $ERV[\triangleleft_c]$ and $ERV[\triangleleft_h]$, exploring heuristics and caching techniques.

\bibliographystyle{splncs04}
\bibliography{references}

\appendix
\chapter*{APPENDIX}
\markboth{APPENDIX}{APPENDIX}

\section{Definitions}

\begin{figure}[h!]    
\centering
    \includegraphics[width=0.6\textwidth]{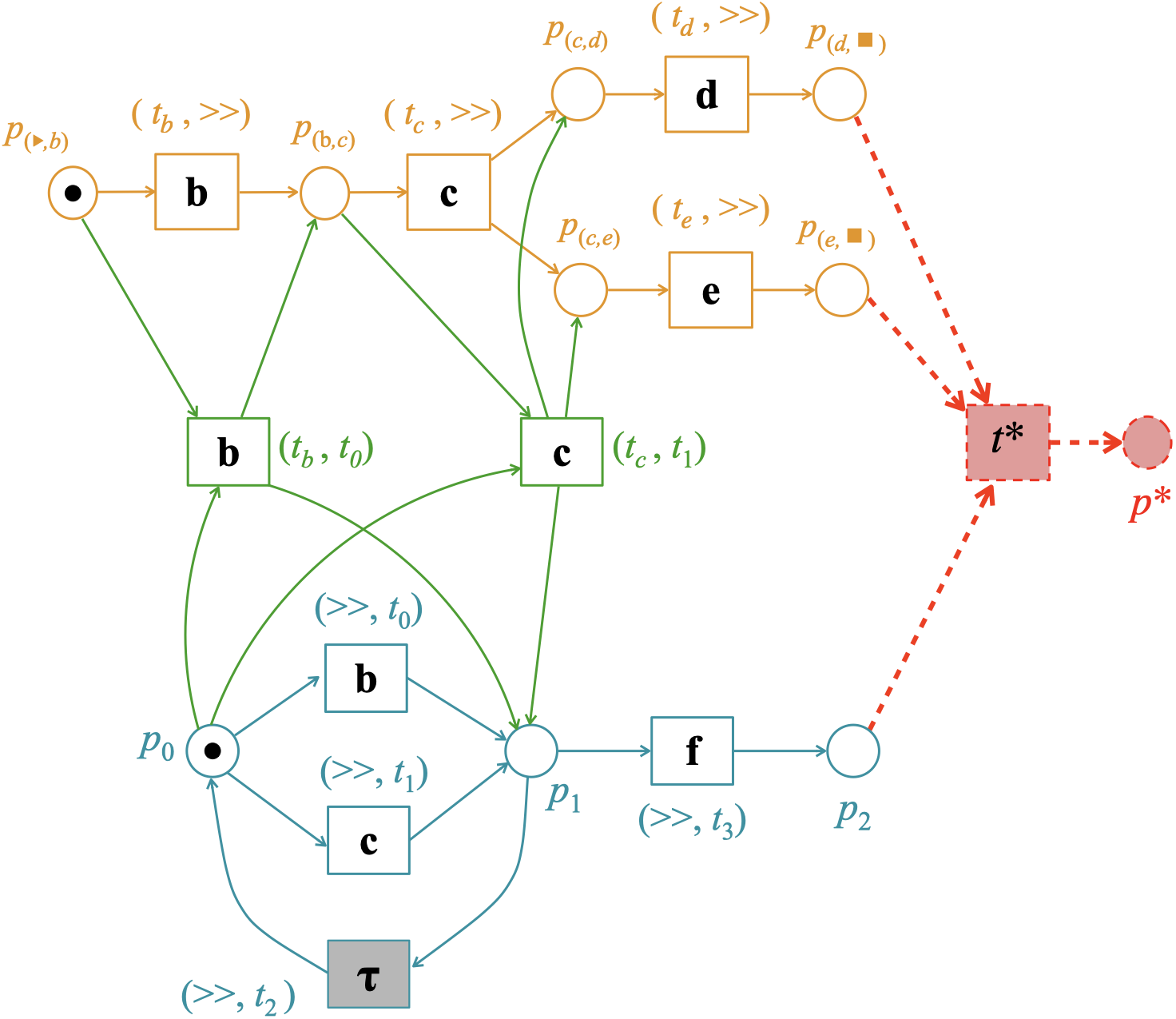}
    \includegraphics[width=0.6\textwidth]{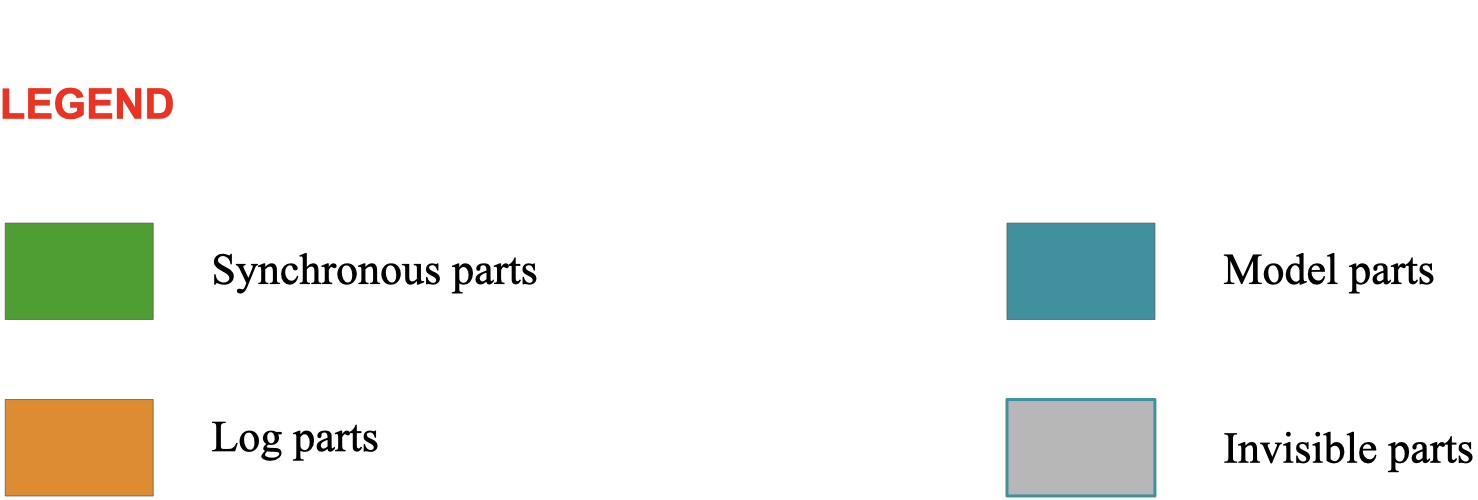}
    \caption{A synchronous product net between a trace net and model net}
    \label{fig:sync-prod-net}
\end{figure}

\Cref{fig:sync-prod-net} shows the synchronous product of two system nets - one originating from the log (trace net) and the other from the model (model net). The parts coloured in \textit{orange} refer to the trace net, while the parts coloured in \textit{blue} represent the process model. The resulting synchronous transitions and arcs are coloured in \textit{green}. Firing a log transition (orange) changes the state of the event net but not the model net, while firing a model transition (blue) changes the state of the model net while keeping the event net's state the same. Firing of a synchronous transition (green) changes the state of both. Formally, a synchronous product net is defined as follows:\\   

% \cite{van2012replaying}
\begin{definition}[Synchronous Product Net of Two System Nets]
    \label{def:sync-prod-net}
    Let there be two system nets $SN_1 = (\mathcal{N}_1, M_{init,1}, M_{final,1}) = ((P_1,T_1,F_1,\lambda_1), \allowbreak M_{init,1}, \allowbreak  M_{final,1})$ and $SN_2 = (\mathcal{N}_2, M_{init,2}, M_{final,2}) = ((P_2,T_2,F_2,\lambda_2), M_{init,2}, \allowbreak M_{final,2})$ over a set of events $E \subseteq \mathcal{E}$ such that $P_1 \cap P_2 = \emptyset$ and $T_1 \cap T_2 = \emptyset$. Then, the \textit{synchronous product net (SPN)} of $\mathcal{N}_1$ and $\mathcal{N}_2$ is the system net $\mathcal{N}_3 = \mathcal{N}_1 \otimes \mathcal{N}_2 = (\mathcal{N}_3, M_{init,3}, M_{final,3}) = ((P_3,T_3,F_3,\lambda_3), M_{init,3}, M_{final,3})$ where
    \begin{itemize}
        \item $P_3 = P_1 \cup P_2$,
        \item $T_3 = (T_1 \times \{\gg\}) \cup (\{\gg\} \times T_2) \cup \{(t_1,t_2) \in (T_1 \times T_2) \ \mid \ \lambda_1(t_1) = \lambda_2(t_2) \neq \uptau\}$,
        \item $W_3 : (P_3 \times T_3) \cup (T_3 \times P_3) \rightarrow \{0,1\}$ is the arc weight function such that 
        \begin{itemize}
            \item $((p_1, \gg), (t_1, \gg)) \in F_3$ if $(p_1,t_1) \in F_1$ where $p_1 \in P_1 \land t_1 \in T_1$,
            \item $((t_1, \gg), (p_1, \gg)) \in F_3$ if $(t_1,p_1) \in F_1$ where $p_1 \in P_1 \land t_1 \in T_1$,
            \item $((\gg, p_2), (\gg, t_2)) \in F_3$ if $(p_2,t_2) \in F_2$ where $p_2 \in P_2 \land t_2 \in T_2$,
            \item $((\gg, t_2), (\gg, p_2)) \in F_3$ if $(t_2,p_2) \in F_2$ where $p_2 \in P_2 \land t_2 \in T_2$,
            \item $((p_1, \gg), (t_1, t_2)) \in F_3$ if $(p_1,t_1) \in F_1$ where $ p_1 \in P_1 \land (t_1,t_2) \in T_3 \cap (T_1 \times T_2)$,
            \item $((\gg, p_2), (t_1, t_2)) \in F_3$ if $(p_2,t_2) \in F_2$ where $p_2 \in P_2 \land (t_1,t_2) \in T_3 \cap (T_1 \times T_2)$,
            \item $((t_1, t_2), (p_1, \gg)) \in F_3$ if $(t_1, p_1) \in F_1$ where $p_1 \in P_1 \land (t_1,t_2) \in T_3 \cap (T_1 \times T_2)$,
            \item $((t_1, t_2), (\gg, p_2)) \in F_3$ if $(t_2, p_2) \in F_2$ where $p_2 \in P_2 \land (t_1,t_2) \in T_3 \cap (T_1 \times T_2)$.
        \end{itemize}
        \item $\lambda_3: T_3 \rightarrow \mathcal{L}^{\uptau}$ is the labelling function, such that for all $(t_1,t_2) \in T_3, \lambda_3((t_1,t_2)) = \lambda_1(t_1)$ if $t_2=\gg$, $\lambda_3((t_1,t_2)) = \lambda_2(t_2)$ if $t_1=\gg$, and $\lambda_3((t_1,t_2)) = \lambda_1(t_1)$ otherwise,  
        \item $M_{init,3} = [M_{init,1}, M_{init,2}]$ is the initial marking,
        \item $M_{final,3} = [M_{final,1}, M_{final,2}]$ is the final marking.
    \end{itemize}
\end{definition}

\section{Proofs}

\begin{theorem}[Adequacy of $\triangleleft_c$]
\label{adequacy}
Let $\upbeta$ be the unfolding of a synchronous product net. If $lc$ is the likelihood cost function on finite configurations, $\triangleleft_c$ is an adequate order on the finite configurations of $\upbeta$.
\begin{proof}
    Let there be two finite configurations $C_1$ and $C_2$ such that $Mark(C_1)=Mark(C_2)$ and $C_1 \triangleleft_c C_2$.
For the first property of adequate orders—well-foundedness, we observe that the values of both $lc(C)$ and $|C|$ are in $\mathbb{N}_0$, which is already a well-founded domain with respect to $<$. Moreover, $>>$ is already a well-founded relation by definition. Hence, relation $\triangleleft_c$ iff $lc(C_1) < lc(C_2) \vee (lc(C_1)=lc(C_2) \wedge |C_1| < |C_2|) \vee (lc(C_1)=lc(C_2) \wedge |C_1| = |C_2| \wedge \phi(C_1) >> \phi(C_2))$ will also be well-founded.

The second property states that $C1 \subset C_2 \implies C_1 \triangleleft_c C_2$. This also holds because $C1 \subset C_2 \implies lc(C_1) \leq lc(C_2)$ since values of $lc$ are in the non-negative domain. In case $lc(C_1)<lc(C_2)$, it naturally follows from definition that $C_1 \triangleleft_c C_2$. In the other case where $lc(C_1)=lc(C_2)$, it still holds that $|C_1|<|C_2|$ and $\phi(C_1) >> \phi(C_2)$ if $C1 \subset C_2$, implying $C_1 \triangleleft_c C_2$. 

Finally, if $E_1$ and $E_2$ are isomorphic extensions to $C_1$ and $C_2$ respectively, observe that they must also preserve homomorphism between the branching process and original system net (recall that two extensions are isomorphic if the labelled (by the homomorphism $h$) directed graphs induced by its events and conditions are isomorphic). This means that $\sum_{e_1 \in E_1} cost(h(e_1)) = \sum_{e_2 \in E_2} cost(h(e_2))$ and hence, $lc(E_1)=lc(E_2)$. Further, for every finite extension $E$ on $C$, $lc(C \oplus E)=lc(C)+lc(E)$. To prove that $C_1 \oplus E_1 \triangleleft_c C_2 \oplus E_2$, let us consider all three cases independently if $C_1 \triangleleft_c C_2$; where either $lc(C_1) < lc(C_2)$ or where $lc(C_1) = lc(C_2)$ but $|C_1| < |C_2|$ or where $lc(C_1)=lc(C_2)$ and $|C_1| = |C_2|$ but $\phi(C_1)>>\phi(C_2)$ :
\begin{enumerate}
    \item if $lc(C_1) < lc(C_2)$, 
    \begin{align*}
        &lc(C_1) + lc(E_1) < lc(C_2) + lc(E_2) \quad (\because lc(E_1)=lc(E_2)) \\
        &\implies lc(C_1 \oplus E_1) < lc(C_2 \oplus E_2) \\
        &\implies C_1 \oplus E_1 \triangleleft_c C_2 \oplus E_2. \\
    \end{align*}
    
    \item if $(lc(C_1) = lc(C_2)) \wedge (|C_1| < |C_2|)$,
    \begin{align*}
        &(lc(C_1) + lc(E_1) = lc(C_2) + lc(E_2)) \wedge (|C_1| + |E_1| < |C_2| + |E_2|)\\
        & \implies lc(C_1 \oplus E_1) = lc(C_2 \oplus E_2) \wedge (|C_1 \oplus E_1| < |C_2 \oplus E_2|) \\
        &\implies C_1 \oplus E_1 \triangleleft_c C_2 \oplus E_2. \\
    \end{align*}
    \item if $(lc(C_1) = lc(C_2)) \wedge (|C_1| = |C_2|) \wedge (\phi(C_1)>>\phi(C_2))$,\\
    
    From case 2., it is easy to see that $lc(C_1 \oplus E_1) = lc(C_2 \oplus E_2) \wedge |C_1 \oplus E_1| = |C_2 \oplus E_2|$ holds.
    If $\phi(C_1)>>\phi(C_2)$, then by properties of the lexigographic order $\phi(C_1 \oplus E_1)>>\phi(C_2 \oplus E_2)$. Therefore, $C_1 \oplus E_1 \triangleleft_c C_2 \oplus E_2$. \\
    
\end{enumerate}
\end{proof}
\end{theorem}

\end{document}